\documentclass[a4paper,10pt]{article}

\pdfoutput=1

 \usepackage[cp1251]{inputenc}

\usepackage{cite,amsmath,amsfonts,amsthm,fullpage}
\usepackage{youngtab}

\usepackage{graphics}
\usepackage{miniplot}
\usepackage{subfigure}

\newcommand{\Pf}{\mathop\mathrm{Pf}\nolimits}

\newcommand{\Pa}{\mathop\mathrm{P}\nolimits}

\newcommand{\DP}{\mathop\mathrm{DP}\nolimits}
\newcommand{\OP}{\mathop\mathrm{OP}\nolimits}

\newcommand{\ttr}{\texttt{tr}}

\newcommand{\f}{\texttt{f}}
\newcommand{\e}{\texttt{e}}
\newcommand{\V}{\texttt{v}}

\newcommand{\Dr}{{\cal {L}}}

\theoremstyle{plain}

\newtheorem{Theorem}{Theorem}
\newtheorem{Lemma}{Lemma}
\newtheorem{Proposition}{Proposition}
\newtheorem{Corollary}{Corollary}
\newtheorem{Remark}{Remark}
\newtheorem{Example}{Example}

\theoremstyle{remark}

\def\ttr{\mathrm {tr}}
\def\det{\mathrm {det}}

\def\diag{\mathrm {diag}}

\def\bp{\begin{Proposition}}
\def\ep{\end{Proposition}}
\def\bc{\begin{Corollary}}
\def\ec{\end{Corollary}}
\def\bl{\begin{Lemma}}
\def\el{\end{Lemma}}
\def\be{\begin{equation}}
\def\ee{\end{equation}}
\def\br{\begin{Remark}\rm\small}
\def\er{\end{Remark}}
\def\brs{\begin{remarks}.\\ \rm\
\begin{enumerate}}
\def\ers{\end{enumerate}\end{remarks}}
\def\bea{\begin{eqnarray}}
\def\eea{\end{eqnarray}}

\def\bx{\begin{Example}\rm\small}
\def\ex{\end{Example}}

\def\det{\mathrm {det}}

\def\diag{\mathrm {diag}}

\def\&{&{\hskip -20pt}}

\newcount\YDcount\YDcount=0
\def\YDsize{10pt}

\def\YD#1{%
\ifnum#1=0
 \ifnum\YDcount=0 \ifx\varnothing\undefined\emptyset\else\varnothing\fi
 \else\vskip1.4pt\egroup\YDcount=0\fi
\else
 \ifnum\YDcount=0 \YDcount=1\vcenter\bgroup\vskip1pt
 \else\nointerlineskip\fi
 \vbox{\hrule\hbox{\vrule height\YDsize
 \loop\hskip\YDsize\vrule\ifnum\YDcount<#1\advance\YDcount1\repeat}\hrule
 \kern-0.4pt}\expandafter\YD
\fi}

\usepackage[usenames,dvipsnames]{color}
\usepackage{ulem}

\def\pb{\mathbf{p}}

\def\xb{\mathbf{x}}
\def\yb{\mathbf{y}}

\def\Zb{\mathbf{Z}}

\def\rB{\rm B}
\def\i{\in\Zb}

\begin{document}

\author{Aleksander Yu.
Orlov\thanks{Institute of Oceanology, Nahimovskii Prospekt 36,
Moscow 117997, Russia, email: orlovs@ocean.ru
}
}
\title{New solvable matrix models III}

\date{December 30, 2021}
\maketitle

\begin{abstract}
We present a family of matrix models whose perturbation series in the coupling constants
are written as the series in projective Schur functions over strict partitions and are examples
of the tau functions of the two-component BKP hierarchy. The case of the usage of the two-component KP
hierarchy is also reviewed and generalized.
\end{abstract}

\bigskip
\medskip

{To Andrey D. Mironov on his 60's birthday}


\section{Introduction \label{Introduction}}

This note was initiated by the work on the generalized Kontsevich model \cite{MMq} and further discussions
with the authors. The perturbation theory series for 
the partition function of this model was written in a very compact form as a sum over strict partitions of 
a pair of projective Schur functions. It was followed by the work \cite{Alex} where the similar series appeared
for the different model (BGW model). The sources of interest to series in projective Schur functions can be
found in 
\cite{Lee2018},\cite{Serbian},\cite{MMN2019},\cite {Alex},\cite{MMNO},\cite{Alex2},\cite{MM-genQ},\cite{MMZ},\cite{MMZ-2},
\cite{MMMZ}. There are some earlier works on the series, 
see \cite{TW},\cite{Or},\cite{ONimmo},\cite{Sho},\cite{LO-2008},\cite{HLO}. On the projective Schur 
functions and the reperesentation theory of the supersymmetric group $q(N)$ and the symmetric group $S_n$,
see \cite{Serg},\cite{Ivanov},\cite{Iv},\cite{EOP},\cite{Strembridge}. 
 The appearence of these functions in the integrable
models was presented in \cite{You},\cite{Nim}.

If a matrix integral is a tau function as a function of its coupling constants I call it solvable. 
As far as I know the first solvable (in this sense) matrix model was presented in the preprint of \cite{GMMMO}
and other examples in \cite{KMMOZ} and \cite{GKM},\cite{KMMM},\cite{Kharchev}, and see the dissertstion \cite{Mir_thesis}
of Andrey Mironov. Then we should point  out the work \cite{L1}.
I marked the family as the third one (``New solvable matrix models III'') because it is 
the direct continuation of \cite{O-2004-New} and also
of \cite{AOV} where the solvable cases were selected.

Here I present and compare two families of solvable matrix integrals.
The first one is a slight generalization of earlier works 
\cite{O-2004-New},\cite{HO2005},\cite{ShiotoOrlov} and in this sense it is not so much new. It is
related to the KP hierarchy of integrable equations. 
The second family is completely new and is related to the KP hierarchy on the root system B (The BKP hierarchy
which was introduced in \cite{DJKM1}).

\paragraph{Notations} We suppose that the reader is familiar with the notions of partitions (we shall use Greek symbols
$\lambda$, $\mu$ for them), strict partitions ( $\alpha,\beta,\gamma,\delta$),
the Schur functions $s_\lambda$, the projective Schur functions $Q_\alpha$.
We denote the set of all partitions by $\Pa$, the set of strict partition $\DP$.
The definitions can be found in \cite{Mac} or, in a brief way in the Appendix.
The details about KP and BKP hierarchies may be found in \cite{JM},\cite{DJKM1} and \cite{KvdLbispec},
see also the textbooks \cite{MJD}, \cite{HB}.

The Schur function can be written  either as the symmetric polynomial in eigenvalues of
a matix $X$, in this case we write it as $s_\lambda(X)$ or as the polynomial in power sum variables
$\pb=(p_1,p_2,\dots)$, then we write it as $s_\lambda(\pb)$. Then $s_\lambda(X)=s_\lambda(\pb(X))$ where
$p_m=\ttr X^m$.

Throughout the text, the matrix size is denoted by $ N $.

The projective function as the polynomial
in power sum variables $\pb_{\rm odd}=(p_1,p_3,\dots)$ is written as $Q_\alpha(\pb_{\rm odd})$. If it is written as
the symmetric polynomial of the eigenvalues of a matrix $X$, then we write it as 
$Q_\alpha(X)=Q_\alpha(\pb_{\rm odd}(X))$, then 
$p_{2m-1}=2\ttr X^{2m-1}$ (pay attention on the factor 2 which is different from the case of the Schur functions
$s_\lambda$).
The same notations will be applied to tau functions related to the series in the Schur functions
(which are 2KP tau functions) and to the series in the projective Schur functions (these are 2BKP 
tau functions).

Consider an (infinite) matrix $A=\{A_{i,j},\,i.j\ge 0  \}$.
For a partitions $\lambda=(\lambda_1,\dots,\lambda_N)$ and $\mu=(\mu_1,\dots,\mu_N)$ whose length
do not exceed $N$ (and can be less than $N$) we introduce
the following notation for the determinant of the $N\times N$ submatrix selected by the 
partitions $\lambda $ and $\mu$:
\be\label{A_l_m(N)}
A_{\lambda,\mu}(N)=\det\left[ A_{N+\lambda_i-i,N+\mu_j-j} \right]_{i,j=1,\dots,N},\quad \lambda,\mu\in\Pa
\ee
In what follows, $N$ is the matrix size and we suppose $N\ge 1$.

Next, for a given pair of strict partitions
$\alpha=(\alpha_1,\dots,\alpha_k)$ and
$\beta=(\beta_1,\dots,\beta_k)$ we shall use the notation
 for the determinant of the submatrix with entries selected by $\alpha$ and $\beta$:
 \be\label{A_alpha_beta}
 A_{\alpha,\beta}:=\det\left[ A_{\alpha_i,\beta_j} \right]_{i,j=1,\dots, k} .\quad \alpha,\beta\in\DP
 \ee

We use notations
$$
\Delta(\xb)=\prod_{i<j\le N}(x_i-x_j),\quad \Delta^*(\xb)=\prod_{i<j\le N}\frac{x_i-x_j}{x_i+x_j}
$$

\paragraph{Hypergeometric tau functions and determinantal formulas.} 
Here we follow \cite{OS-TMP}. Such tau functions appeared
in \cite{GKM} in a different form.
Let $r$ be a function on the lattice $\mathbb{Z}$.
Consider the following series
\be\label{tau-elementary}
 1+r(n+1)x+r(n+1)r(n+2)x^2+r(n+1)r(n+2)r(n+3)x^3 + \cdots =: \tau_r(n;x,1)
\ee
Here $n$ is an arbitrary integer.
This is just a Taylor series for a function $\tau_r$ with a given $n$  written in a form that is as close as 
possible to typical 
hypergeometric series. Here $n$ is just a parameter that will be of use later. 
If as $r$ we take a rational (or trigonometric) function, we get a generalized (or basic) hypergeometric series.
Indeed, take 
\be\label{r-rational}
r(n)=\frac{\prod_{i=1}^p(a_i+n)}{n\prod_{i=1}^q(b_i+n)}
\ee
We obtain
$$
\tau_r(n;x)=\sum_{m\ge 0} \frac{\prod_{i=1}^p(a_i+n)_m}{m!\prod_{i=1}^q(b_i+n)_m} x^m=
{_{p}F}_q\left({a_1+n,\dots,a_p+n\atop b_1+n,\dots, b_q+n} \mid x\right)
$$
where 
\be
(a)_n=a(a+1)\cdots(a+n-1)=\frac{\Gamma(a+n)}{\Gamma(a)}
\ee
is the Pochhammer symbol.

One can prove the following formula 
which express a certain series over partitions in terms of (\ref{tau-elementary})
(see for instance \cite{OS-TMP})
\be\label{tau(nXY)}
\tau_r(n;X,Y):= \sum_{\lambda} r_\lambda(n) s_\lambda(X)s_\lambda(Y) =
\frac{c_n}{c_{n-N}}\frac{\det \left[ \tau_r(n-N+1;x_iy_j,1) \right]_{i,j\le N}}
{ \Delta(\xb)\Delta(\yb)}
\ee
 where $c_k=\prod_{i=0}^{k-1}\left(r(i)  \right)^{i-k}$  and where
\be\label{content-product}
r_\lambda(n):=\prod_{(i,j)\in\lambda} r(n+j-i),\quad r_{(0)}(n)=1
\ee
where the product ranges over all nodes of the Young diagram $\lambda$
is the so-called content product (which has the meaning of the generalized
Pochhammer symbol related to $\lambda$). Notice that in case $N=1$ we can write $\tau_r(n;x,y)=\tau_r(n;xy,1)$.
Let us test the formula for the simplest case $r\equiv 1$:
\be
\det\left[1-X\otimes Y\right]=\sum_{\lambda} s_\lambda(X)s_\lambda(Y)
=\frac{\det\left[( 1-x_iy_j )^{-1} \right]_{i,j\le N}}
{\Delta(\xb)\Delta(\yb)}
\ee

Sometimes we also use infinite sets of power sums $\pb^{i}=(p^{(i)}_1,p^{(i)}_1,\dots )$ and instead of matrices
$X$ and $Y$ set:
\be\label{tau-p1-p2}
\tau_r(n;\pb^1,\pb^2):= \sum_{\lambda} r_\lambda(n) s_\lambda(\pb^1)s_\lambda(\pb^2)
\ee
An example $r\equiv 1$:
$$
\tau_1(n;\pb^1,\pb^2):=e^{\sum_{m>0} \frac 1m p_m^{(1)} p_m^{(2)} }
$$
In case the function $r$ has zeroes, there exists a determinantal representation. 
Suppose $r(0)$, then $r_\lambda(n)=0$ if $\ell(\lambda)>n$. Then
 \be\label{tau(npp)}
\tau_r(n;\pb^1,\pb^2)= \sum_{\lambda\atop\ell(\lambda)\le n} r_\lambda(n) s_\lambda(\pb^1)s_\lambda(\pb^2)
= c_n \det\left[ \partial_{p_1^{(1)}}^{a}\partial_{p_1^{(2)}}^{b} \tau_r(1;\pb^1,\pb^2) \right]_{a,b=0,\dots,n-1},
\ee
where $c_k=\prod_{i=1}^{k-1}\left(r(i)  \right)^{i-k}$, and where
$$
\tau_r(1;\pb^1,\pb^2)=1+\sum_{m>0}r(1)\cdots r(m)s_{(m)}(\pb^1)s_{(m)}(\pb^2),
$$
 see \cite{O-2004-New}. 
 
 In addition there is 
the following formula \cite{OS2000}:
\be\label{tau(nXp)}
\tau_r(n;X,\pb)= \sum_{\lambda} r_\lambda(n) s_\lambda(X)s_\lambda(\pb)=
\frac{\det\left[ x_i^{N-k}
\tau_r\left(n-k+1; x_i,\pb\right) \right]_{i,k\le N}}
{\det \left[ x_i^{N-k} \right]}
\ee
where
\be
\tau_r\left(n; x_i,\pb\right)=1+\sum_{m > 0}r(n)\cdots r(n+m) x_i^m s_{(m)}(\pb)
\ee

Let us test it for $r\equiv 1$, $\pb=\pb_\infty:=(1,0,0,\dots)$:
\be
e^{\ttr X}= \sum_\lambda s_\lambda(X)s_\lambda(\pb_\infty)=
\frac{\det\left[ x_i^{N-k}
e^{x_i} \right]_{i,k\le N}}
{\det \left[ x_i^{N-k} \right]}
\ee

There are similar series
$$
\sum_{\lambda} r_\lambda(n) s_\lambda(X)
$$
which can be written as a Pfaffian \cite{OST-I}, however we will not use them in the present text.

\section{New solvable families of matrix models}

In this subsection general formulas for new solvable families of matrix models are written down 
which will be further applied for certain ensembles of random matrices. Namely new types of coupling
are introduced.

The cases of unitary and of Hermitian matrices are considered in the next sections.

\paragraph{Solvable family related to the two-component KP hierarchy}
The first one is related to the two-component KP hierarchy of integrable equations and it
is just a generalization of a number of well-known matrix integrals
whose partition function is a double series in the Schur functions
over partitions. Actually it is a slight generalization of earlier works 
\cite{O-2004-New},\cite{HO2005},\cite{ShiotoOrlov} and in this sense it is not so much new.
This solvable family could be written as 
follows:
\be\label{2KP_2KP_2KP=2KP}
\int \tau_{A^1}(\pb^1,X^\dag)\tau_r(N;XY,\mathbb{I}_N)\tau_{A^2}(Y^\dag,\pb^2)\det\left(v(X)u(Y) \right) 
d\Omega_N(X,Y)=\tau_{A^3}(\pb^1,\pb^2)
\ee
where the relation
\be\label{composition_law}
A^3=A^1 g^{r,v,u} A^2
\ee
which can be called the {\it composition rule}, is satisfied.

The notations are as follows:

Two infinite sets $\pb^i=(p^{(i)}_1,p^{(i)}_1,\dots)$ are the sets of the coupling constants.
The measure $d\Omega_N(X,Y)$ on the space of $N\times N$ matrices $X$ and $Y$ is defined by the choice of 
matrix ensembles. Ensembles of unitary, of Hermitian, of complex and of normal matrices $X,Y$ should be separately
considered. The determinant $\det(u(X)v(Y))$ (where $u$ and $v$ are given functions)
can be considered part of the measure, but we want to 
highlight it.
The term which pairs the matrices $X$ and $Y$, namely $\tau_r$, which is labeled by a function $r$ on the lattice 
$\Zb$, is the series (\ref{tau(nXY)}) of the following form
\be\label{pairing}
\tau_r(N;XY,I_N) :=\sum_\lambda s_\lambda(XY) s_\lambda(I_N)\prod_{(i,j)\in \lambda} r(N+j-i)
\ee
where $I_N$ is the identity matrix and $s_\lambda(XY)$ is  the Schur function 
indexed by a partition $\lambda$. The sum ranges over all partitions and actually is restricted
by the condition that the number of parts of each $\lambda$ (called the length $\ell(\lambda)$ of $\lambda$)
does not exceed $N$ because otherwise $s_\lambda$ vanishes. The product in the right hand side over nodes
with coordinates $i,j$
of the Young diagram of $\lambda$ is called
content product\footnote{It is of use in the representation theory of the symmetric group, in the context 
of soliton theory, see a short review in \cite{H-review}.}.
The main examples of the $X$-$Y$ pairing given by (\ref{pairing}) are
$$
e^{c\ttr XY} = \sum_{\lambda} s_\lambda(XY)s_\lambda(\pb_\infty) c^{|\lambda|},\quad 
\pb_\infty=(1,0,0,\dots)
$$
$$
\det(1-cXY)^{-a}=
\sum_{\lambda} s_\lambda(XY)s_\lambda(\pb_\infty) c^{|\lambda|}\prod_{(i,j)\in\lambda}(a+j-i),
$$
$$
{_pF}_q\left ({{N+a_1,\dots, N+a_p}\atop{N+b_1,\dots, N+b_q}}\mid XY\right)=
\sum_{\lambda}\frac{\prod_{i=1}^p (N+a_i)_\lambda}{\prod_{i=1}^q (N+b_i)_\lambda} s_\lambda(XY)s_\lambda(\pb_\infty)
$$
related resectively to $r(x)=cx^{-1}$, $r(x)=c\frac{x+a}{x}$ and to 
$r(x)=\frac{(a_1+x)\cdots (a_p+x)}{(b_1+x)\cdots(b_q+x)}$ and ${_pF}_q$ is the hypergeometric function of matrix 
argument \cite{VK}.
In (\ref{2KP_2KP_2KP=2KP})
$A^1,A^2,A^3$ are three infinite matrices which define respectively tau functions $\tau_{A^1},\tau_{A^2},\tau_{A^3}$
which are tau functions of the two-component KP hierarchy and can be written in form
\be\label{tauA1}
\tau_{A^1}(N;\pb^1,X)=\sum_{\lambda,\mu} s_\lambda(\pb^1)s_\mu(X)A^1_{\lambda,\mu}(N)
\ee
\be\label{tauA2}
\tau_{A^2}(N;\pb^2,Y)=\sum_{\lambda,\mu} s_\lambda(\pb^2)s_\mu(Y)A^2_{\lambda,\mu}(N)
\ee
\be\label{tauA3}
\tau_{A^3}(N;\pb^1,\pb^2)=\sum_{\lambda,\mu} s_\lambda(\pb^1)s_\mu(\pb^2)A^3_{\lambda,\mu}(N)
\ee
where $A^i_{\lambda,\mu}(N)$ are defined as in (\ref{A_l_m(N)}),
see (\ref{tau_ferm}) for the fermionic representation of such two-component KP tau function.
In what  follows if possible I will omit the argument $N$.

\br
In the case when each of the matrices $X$, $Y$ belongs to one of the groups $\mathbb{U}_N$, $\mathbb{GL}_N(C)$ 
and when, in addition to this, both matrices $A^1$, $A^2$ are diagonal, such a model was presented 
in \cite{AOV} and is related to the graph (c) resembling the number 8 , see Fig 1  below.
\er

\br
In the Appendix we present examples of tau function (\ref{tauA1}), (\ref{tauA2}) obtained as
$n_1$-matrix integrals over unitary and complex matrices from the works \cite{AOV},\cite{NO2020tmp}.
In this case tau function (\ref{tauA3}) considered in the right hand side of (\ref{2KP_2KP_2KP=2KP}) below
is equal to a $(n_1+n_2+2)$-matrix integral.
\er

The matrix $A^3$ is defined by the choice of $A^1$ and $A^2$ and is equal to the product of three (infinite) matrices,
see (\ref{composition_law})
where $g^{r,u,v}$ is the matrix of moments:
\be\label{moments-gen}
 g^{r,u,v}_{i,j}=K_N\int \int \bar{x}^i\bar{y}^j \tau_r(1;xy,1)v(x)u(y) d\Omega_1(x,y)
 \ee
 where the bar denotes the complex conjugation and where
 \be\label{K_N}
 K_N=\frac{1}{\prod_{k=1}^{N-1}\prod_{i=1}^{k} r(i)}
 \ee
Here $d\Omega_1(x,y)$ is $d\Omega_N(X,Y)$ taken for $N=1$ and 
\be\label{tau_for_moments}
\tau_r(1;xy,1)=\sum_{m\ge 0}x^my^m r(1)\cdots r(m)
\ee
see (\ref{pairing}). The integration measure is defined by the choice of matrix ensemble.

The main example of self-coupling terms $\tau_{A^1}$ and $\tau_{A^2}$ is the case 
$A^1=A^2=\mathbb{I}_\infty$ (identity matrix):
\be\label{tau_simplest}
\tau_{A^1}(\pb^1,X)= e^{\sum_{m} \frac 1m p^{(1)}_m\ttr X^m},\quad 
\tau_{A^2}(\pb^2,Y)= e^{\sum_{m} \frac 1m p^{(2)}_m\ttr Y^m}
\ee

The tau functions $\tau_{A^1}$ and $\tau_{A^2}$  which depend on the choice of $A^{i}$ and of $\pb^i$, $i=1,2$
dessribe the self-coupling of matrices $X$ and $Y$ respectively. 
Tau function $\tau_r$ describes the coupling between $X$ and $Y$ and depend on the choice of the function $r$.

\paragraph{Solvable models related to the two-component BKP hierarchy}

In this article I am going to show that the simple replacement of $\tau_r(N;XY,\mathbb{I}_N)$ by 
$\tau_r(N;X^2Y^2,\mathbb{I_N})$
and the replacements of 2KP tau functions $\tau_{A^i}$, $i=1,2,3$ by 2BKP tau functions $\tau^{\rB}_{A^i}$: 
\be\label{tauB1}
\tau^{\rB}_{A^1}(\pb^1_{\rm odd},X^\dag)=
\sum_{\alpha,\beta\in\DP} 2^{-\ell(\alpha)}Q_\alpha(\pb^1_{\rm odd})Q_\beta(X^\dag)A^1_{\alpha,\beta}
\ee
\be\label{tauB2}
\tau^{\rB}_{A^2}(\pb^2_{\rm odd},Y^\dag)=
\sum_{\alpha,\beta\in\DP} 2^{-\ell(\alpha)}Q_\alpha(\pb^2_{\rm odd})Q_\beta(Y^\dag)A^2_{\alpha,\beta}
\ee
\be\label{tauB3}
\tau^{\rB}_{A^3}(\pb^1_{\rm odd},\pb^2_{\rm odd})=
\sum_{\alpha,\beta\in\DP} 2^{-\ell(\alpha)}Q_\alpha(\pb^1_{\rm odd})Q_\beta(\pb^2_{\rm odd})A^3_{\alpha,\beta}
\ee
where $A^a_{\alpha,\beta}$ are defined by (\ref{A_alpha_beta}) applied to $A=A^a,\,a=1,2,3$.
See (\ref{tau_ferm}) for the fermionic representation of such two-component KP tau function.  
Let us presents another family of matrix models:
\be\label{2BKP_2KP_2BKP=2BKP}
\int \tau^{\rB}_{A^1}(\pb^1_{\rm odd},X^\dag)
\tau_r(N;X^2Y^2,\mathbb{I}_N)\tau^{\rB}_{A^2}(Y^\dag,\pb^2_{\rm odd})
\det\left(v(X)u(Y)  \right)   d\Omega_N(X,Y)
\ee
\be
=\tau^{\rB}_{A^3}(\pb^1_{\rm odd},\pb^2_{\rm odd})
\ee
with the same composition rule (\ref{composition_law}) 
with the replacement of $\tau_r(1;xy,1)$ by 
$\tau_r(1;x^2y^2,1)$ in (\ref{moments-gen}) and
where now $\pb^i=(p^{(i)}_1,p^{(i)}_3,p^{(i)}_5,\dots)$, 
$\tau^{\rB}_{A^i}$, $i=1,2$ is a pair of 2BKP tau functions, $\tau_r$ is given by (\ref{pairing}) and 
$\tau^{\rB}_{A^3}(\pb^1_{\rm odd},\pb^2_{\rm odd})$ is the 2BKP tau function which is written as the double series 
over strict partitions in projective Schur functions with explicitly written coefficients.
Now we have
\be\label{moments-genB}
 g^{r,u,v}_{i,j}=K_N\int \int \bar{x}^i\bar{y}^j \tau_r(1;x^2y^2,1)v(x)u(y) d\Omega_1(x,y)
 \ee
where
 \be\label{tau_for_momentsB}
\tau_r(1;x^2y^2,1)=\sum_{m\ge 0}x^{2m}y^{2m} r(1)\cdots r(m)
\ee
instead of (\ref{moments-gen}),(\ref{tau_for_moments}).

It can be written also as fermonic vacuum expectation value, see (\ref{tauB_ferm}), and can be presented 
in the pfaffian form.
We will consider few different matrix ensembles, each type of matrices defines the choice of the 
integration measure $d\Omega_N$.

One of the differences between families (\ref{2BKP_2KP_2BKP=2BKP}) and (\ref{2KP_2KP_2KP=2KP}) is
that in the last case the ensembles of complex and of normal matrices are included, and it is not possible
to do in version (\ref{2BKP_2KP_2BKP=2BKP}).

The examples of self-coupling terms $\tau^{\rB}_{1,2}$ are
\be\label{tau_B_simplest}
\tau^{\rB}_{A^1}(\pb^1_{\rm odd},X)=e^{\sum_{m>0,{\rm odd}} \frac1m p^{(1)}_m\ttr X^m},\quad 
\tau^{\rB}_{A^2}(\pb^2_{\rm odd},Y)=e^{\sum_{m>0,{\rm odd}} \frac1m p^{(2)}_m\ttr Y^m}
\ee
compare to (\ref{tau_simplest}).

The examples of the $X$-$Y$ pairing are
\be\label{pairingBKP}
e^{c\ttr X^2Y^2},\quad \det(1-cX^2Y^2)^{-a},
\quad {_pF}_q\left ({{N+a_1,\dots, N+a_p}\atop{N+b_1,\dots,N+ b_q}}\mid X^2Y^2\right) 
\ee

\br
Instead of tau function of two-component tau function $\tau^{\rm B}_{A^i}(X,\pb^i)$ (where $i$ is either 1 or 2) 
one can consider a one component
tau function in form $\sum_\alpha Q_\alpha(X)B_\alpha$, where $B_\alpha$ is a pfaffian (for instance, see 
\cite{OST-I}) and have the meaning of the ``Cartan coordinate'' of the related point in the isotropic grassmannian 
\cite{HB}. Then in the right hand side of 
(\ref{2BKP_2KP_2BKP=2BKP}) we get a certain one-component BKP 
tau function. The similar remark is true for the right hand side of (\ref{2KP_2KP_2KP=2KP}) where ones can
take a certain Sato series $\sum_\alpha s_\lambda(X)\pi_\lambda$ (where $\pi_\lambda$ is the ``Plucker coordinate'')
instead of $\tau_{A^i}(X,\pb^i)$ in the left hand side.
\er

\br\label{gauge_freedom}
Thus, we have three group of parameters to construct the families (\ref{2KP_2KP_2KP=2KP}) and (\ref{2BKP_2KP_2BKP=2BKP}),
these are matrices $A^1,A^2$ and functions $r,v,u$. We have the transformation group 
$A^1\to A^1S_1,\,A^2\to S^{-1}_2A^2,\,g^{r,v,u}\to S^{-1}_1g^{r,v,u}S_2$ to get equivalent models as we see from 
(\ref{composition_law}). The choice of possible $g^{r,v,u}$ is rather restriced by the type of measure in 
(\ref{moments-gen}) and in (\ref{moments-genB}).
\er

\br
Notice that zeroes of $r$ do not allow to consider (\ref{tau_for_momentsB}),(\ref{L2eq}) as a 
particular case of (\ref{tau_for_moments}),(\ref{L1eq}).
\er

\paragraph{Chain matrix models}

The use of tau functions in both sides of (\ref{2KP_2KP_2KP=2KP}) and also of (\ref{2BKP_2KP_2BKP=2BKP})  
allows to consider multimatrix models
since tau functions under the integral can be treated as multimatrix integrals themselves.
The chain ensemble of $2n$ matrices is obtained as follows:

In the 2KP case:

\be\label{chain_A}
\int \tau_{A^1}(\pb^1,X^\dag_1)K(X_1,\dots,X_{2n-1})
\tau_{r^{(2n-1)}}(N;X_{2n-1}X_{X_{2n}})\tau_{A^{2n}}(X^\dag_{2n},\pb^2)d\Omega
=\tau_{A^{2n+1}}(\pb^1,\pb^2)
\ee
where
$$
K(X_1,\dots,X_{2n-1})=
\prod_{i=1}^{n-1}\tau_{r^{(i)}}(N;X_{2i-1}X_{2i})\tau_{A^{i+1}}(X^\dag_{2i},X^\dag_{2i+1})
$$
(we put $K=1$ for $n=1$) and
$$
d\Omega =\prod_{i ,\, i\,{\rm odd}}^{2n-1}\det\left(v(X_i)u(X_{i+1}) \right) d\Omega_N(X_i.X_{i+1})
$$
The composition rule is
\be\label{chain_composition}
A^{2n+1}=A^1 g^{r^{(1)},v^{(1)},u^{(1)}}A^{2}g^{r^{(2)},v^{(2)},u^{(2)}}\cdots 
A^{2n-1} g^{r^{(2n-1)},v^{(2n-1)},u^{(2n-1)}}A^{2n}
\ee

This chain model contains the chain models considered in \cite{AOV} and in \cite{O-2004-New},\cite{HO2005} (Appendix).

\br
In the appendix we present

\er

In the 2BKP case is obtained by the replacement of each $\tau_{r^{(i)}}(N;X_{2i-1}X_{X_{2i}})$ 
by $\tau_{r^{(i)}}(N;X^2_{2i-1}X^2_{X_{2i}})$:

\be\label{chain_B}
\int \tau^{\rB}_{A^1}(\pb^1_{\rm odd},X^\dag_1)K(X_1,\dots,X_{2n-1})
\tau_{r^{(2n-1)}}(N;X^2_{2n-1}X^2_{X_{2n}})\tau^{\rB}_{A^{2n}}(X^\dag_{2n},\pb^2_{\rm odd})d\Omega
=\tau^{\rB}_{A^{2n+1}}(\pb^1_{\rm odd},\pb^2_{\rm odd})
\ee
where $d\Omega$ is the same as in the previous case, where
$$
K(X_1,\dots,X_{2n-1})=
\prod_{i=1}^{n-1}\tau_{r^{(i)}}(N;X^2_{2i-1}X^2_{2i})\tau^{\rB}_{A^{i+1}}(X^\dag_{2i},X^\dag_{2i+1})
$$
and the same composition rule (\ref{chain_composition}). 

The pairing of the neighboring matrices 
 $X_{2i}$ and $X_{2i+1}$ is given by the choice of $\tau^{\rB}_{A^{}}(X_{2i},X_{2i+1})$ and the pairing of
 $X_{2i-1}$ and $X_{2i}$ is given by $\tau_{r^{(i)}}(X^2_{2i-1},X^2_{2i})$ of form (\ref{pairing})
 defined by the choice of functions $r^{(1)},r^{(2)},\dots$.
This chain is different from the one discussed in \cite{KMMMP} and from the one discussed in \cite{AOV}.

The mixed chain where both 2KP and 2BKP tau fnctions are used is natural in this context.

 \br\label{shift} As it is well-known, the replacements
\be
\tau(\pb^1,\pb^2)\to e^{\sum_{m>0} \left(a_mp^{(1)}_m +b_mp^{(2)}_m\right)}\tau(\pb^1,\pb^2),\quad 
\tau^{\rm B}(\pb^1,\pb^2)\to e^{\sum_{m>0,{\rm odd}} \left(a_mp^{(1)}_m +b_mp^{(2)}_m\right)  }
\tau^{\rm B}(\pb^1,\pb^2)
\ee
with any sets of parameters $\{a_m\}$ and $\{b_m\}$
convert a tau function to a tau function in both 2KP and 2BKP cases. This is taken into account
by the freedom in the choice of factors $\det \left(v(X)\right)$ and $\det\left( u(Y)\right)$ 
in the integration measure.
\er

\subsection{Technical tools}

There is a number of useful lemmas:
\bl\label{L1}\cite{O-2004-New}
\be\label{L1eq}
\int_{\mathbb{U}_N} \tau_r(UXU^\dag Y,\mathbb{I}_N) d_*U=
K_N\frac{\det\left[\tau_r(x_iy_j,1)  \right]_{i,j} }{\Delta(\xb)\Delta(\yb)}
\ee
where $K_N$ is given by (\ref{K_N})
and where $\tau_r(UXU^\dag Y,\mathbb{I}_N)$ was defined in (\ref{pairing}) and where $\int_{\mathbb{U}_N} d_*U=1$.
\el

\bl\label{L2}\cite{O-2004-New}
\be\label{L2eq}
\int_{\mathbb{GL}_N(C)} \tau_r(N;ZXZ^\dag Y,\pb_\infty) d\Omega(Z,Z^\dag)=
K_N\frac{\det\left[\tau_r(x_iy_j,1)  \right]_{i,j} }{\Delta(\xb)\Delta(\yb)}
\ee
where
\be
d\Omega_N(Z,Z^\dag)=C_N e^{-\ttr ZZ^\dag}\prod_{}d\Re Z_{i,j}d\Im Z_{i,j},
\ee
where $\tau_r(N;ZXZ^\dag Y,\pb_\infty)$ was defined in (\ref{pairingZ}) and 
$\int_{\mathbb{GL}_N(C)} d\Omega_N(Z,Z^\dag)=1$.

\el

\br\label{main_idea}
A very simple technical idea to go from (\ref{2KP_2KP_2KP=2KP}) to (\ref{2BKP_2KP_2BKP=2BKP})
is to use the fact that if we replace the sets $\{x_i,y_i\}$ by $\{x^2_i,y^2_i\}$ then
the result of the multiplication of the both left hand sides
in (\ref{L1eq}) and (\ref{L2eq}) by the integration measure of certain matrix ensembles $d\Omega_N$ 
results in the factor $\frac{\left(\Delta(\xb)\Delta(\yb)\right)^2}{\Delta(x^2)\Delta(y^2)}=\Delta^*(\xb)\Delta^*(\yb)$ which is typical for the 
multifold integrals related to BKP, see Appendix in \cite{Or} and \cite{HLO} and \cite{OST-II}.
These ensembles are ensemble of two unitary matrices, the ensemble of two Hermitian matrices
and some others. However it fails for ensembles of complex and of normal matrices.
\er

Next:
\bl\label{<ss>}
For $\lambda,\mu\in\Pa$ we have
\be
<s_\lambda,s_\mu>^{r,v,u}_{d\Omega}:=
\int s_\lambda(X)s_\mu(Y)\Delta(\xb)\Delta(\yb)\prod_{i=1}^N \tau_r(1;x_iy_i,1)v(x_i)u(y_i)d\Omega_1(x_i,y_i)
=g^{r,v,u}_{\lambda,\mu}
\ee
where
\be\label{g_lambda,mu}
g^{r,v,u}_{\lambda,\mu}(N)=\det\left[g^{r,v,u}_{\lambda_i-i+N,\mu_i-i+N} \right]_{i,j}
\ee
see (\ref{A_l_m(N)}),
and $g^{r,v,u}_{i,j}$ is given by (\ref{moments-gen}).
\el

\bl\label{<QQ>} For $\alpha,\beta\in\DP$:
\be
<Q_\alpha,Q_\beta>^{r,v,u}_{d\Omega}:=
\int Q_\alpha(X)Q_\beta(Y)\Delta^*(\xb)\Delta^*(\yb)\prod_{i=1}^N \tau_r(1;x^2_iy^2_i,1)v(x_i)u(y_i)d\Omega_1(x_i,y_i)
=g^{r,v,u}_{\alpha,\beta}
\ee
where
\be\label{g_alpha,beta}
g^{r,v,u}_{\alpha,\beta}=\det\left[g^{r,v,u}_{\alpha_i,\beta_j} \right]_{i,j}
\ee
and $g^{r,v,u}_{i,j}$ is given by (\ref{moments-genB}).
\el
 In a more detailed text the mixed scalar product $<s_\lambda,Q_\alpha>^{r,v,u}_{d\Omega}$ 
 where we replace $\tau_r(1;x_iy_i,1)$
 by $\tau_r(1;x_iy^2_i,1)$ will be considered.

 At last we have natural
 \bl\label{product}
 Let $A=\{A_{i,j},\,i,j\ge 0\}$ and $B=\{B_{i.j},\,i,j\ge 0\}$ be (infinite) matrices and their product $C=AB$ exists.
 Then
 \be
 C_{\alpha,\beta}:=\sum_{\gamma\in\DP} A_{\alpha,\gamma}B_{\gamma,\beta}
 \ee
 where we use notation (\ref{A_alpha_beta}).
 \el
 This is complete set of tools to prove relations (\ref{2KP_2KP_2KP=2KP}),(\ref{2BKP_2KP_2BKP=2BKP}),
 (\ref{2KP_2KP_2KPZ=2KP}),(\ref{2BKP_2BKP_2BKPZ=2BKP}),
 (\ref{chain_A}),(\ref{chain_B}) for ensembles of unitary and ensembles of Hermitian matrices.

\subsection{Both matrices are unitary}

\paragraph{A. Preliminary}
In \cite{ZinZub} the following equality was proven
\be\label{ZZ1}
\int_{\mathbb{U}_N\times\mathbb{U}_N} e^{\ttr U_1U_2+\sum_{m>} \frac1m p^{(1)}_m\ttr U^{-m}_1 + 
\sum_{m>} \frac1m p^{(2)}_m\ttr U^{-m}_2}d_*U_1d_*U_2 =\sum_{\lambda} s_\lambda(\pb^1)s_\lambda(\pb^2)
\prod_{(i,j)\in\lambda}\frac{1}{N+j-i}
\ee
where $s_\lambda$ denotes the Schur function \cite{Mac} indexed by a partition $\lambda$.
The seria over partitions in the right hand side is an example of the family called KP hypergeometric 
tau functions,
see \cite{GKM},\cite{OS-TMP} and can be presented as the determinant of the hypergeometric
functions ${}_1F_1$ for a special choice of $\pb^1,\pb^2$. 
\br
In the approach of work \cite{AOV}
this matrix integral is related to the graph (c) on Fig 1 resembling number 8, where we have 3 faces (decorated 
with three 'monodromies' $U_1U_2$, $U^\dag_1$, $U^\dag_2$ under the integral), two ribbon edges (two 'random
matrices' $U_1,U_2$) and one vertex (decorated with the 'monodromy' $\mathbb{I}_N$ in the example).
\er
This is contained in the family (\ref{2KP_2KP_2KP=2KP}) where $X,Y\in\mathbb{U}_N$,
$A^1=A^2=\mathbb{I}_N$, $v=u\equiv 1$ and $r(x)=x^{-1}$. 

Another example is the choice $A^1=A^2=\mathbb{I}_N$ and 
$\tau_r(N;U_1U_2,\mathbb{I}_N)=\det\left(\mathbb{I}_N-cU_1U_2 \right)^{-a}$
as it was suggested in \cite{HO2005} and studied in \cite{Bertola_}.

After some work (see the next paragraph for detais) we obtain
\be\label{J_U}
J^{A^1,A^2,r,v,u}_{\mathbb{U}_N\times\mathbb{U}_N}(\pb^1,\pb^2):=
\ee
$$
\int_{\mathbb{U}_N\times\mathbb{U}_N} \tau_{A^1}(\pb^1,U^\dag_1)\tau_r(N;U_1U_2,\mathbb{I}_N)\tau_{A^2}(U^\dag_2,\pb^2)
\det\left( v(X) u(Y) \right)  d_*U_1 d_*U_2
$$
\be
 =\sum_{\lambda,\mu\in\Pa\atop\ell(\lambda),\ell(\mu)\le N} 
 s_\lambda(\pb^1)s_\mu(\pb^2)\sum_{\tilde{\lambda},\tilde{\mu}\in\Pa}
 A^1_{\lambda,\tilde{\lambda}}(N)g^{r,v,u}_{\tilde{\lambda},\tilde{\mu}}(N)A^2_{\tilde{\mu},\mu}(N)
\ee
where $\tau_{A^i},\,i=1,2$ are given by (\ref{tauA1}),(\ref{tauA2}), $\tau_r(N;U_1U_2,\mathbb{I}_N)$ is given by
(\ref{pairing}), $g^{r,v,u}_{\tilde{\lambda},\tilde{\mu}}(N)$ is given by (\ref{g_lambda,mu}) where
according to (\ref{moments-gen}) we have
$$
g^{r,v,u}_{i,j}=-\frac{1}{2\pi^2}\oint\oint x^{-i}y^{-j}\tau_r(1;xy,1) v(x)u(y) \frac{dx}{x}\frac{dy}{y}
$$
where $\tau_r(1;xy,1)$ is given by (\ref{tau_for_moments}).
The family is parametrized by the choice of $A^i,\,i=1,2$ and by the choice of functions $r,v,u$
and one keeps in mind Remark \ref{gauge_freedom}. Thanks to Lemma \ref{product} the relation 
(\ref{composition_law}) is fulfilled.

Here we present different family of solvable integrals and our main example which is similar to 
(\ref{ZZ1}) is the Example \ref{ZZ2}.
For us it is more suitable to change $U_1,U_2$ respectively to $X,Y$ below.

\paragraph{B. New family.} 
We will study the integral
\be\label{I_U}
I^{A^1,A^2,r,v,u}_{\mathbb{U}_N\times\mathbb{U}_N}(\pb^1_{\rm odd},\pb^2_{\rm odd})=
\ee
$$
\int_{\mathbb{U}_N\times\mathbb{U}_N} 
\tau_{A^1}(\pb^1_{\rm odd},X^\dag)\tau_r(N;X^2Y^2,\mathbb{I}_N)\tau_{A^2}(Y^\dag,\pb^2_{\rm odd})
\det\left( v(X) u(Y) \right)  d\Omega_N(X,Y)
$$
where $d\Omega_N(X,Y)$ is the product of Haar measures $d_*X d_*Y$
on $\mathbb{U}_N$. Using the standard separation of variables in $X,Y\in\mathbb{U}_N$ see \cite{Mehta}
 $X=U_1\diag(x_i)U^{-1}_1$, $Y=U_2\diag(y_i)U^{-1}_2$,
where $U_1,U_2\in \mathbb{U}_N$, we can write
$$
d_*X d_*Y=|\Delta(\xb)\Delta(\yb)|^2 \prod_{i=1}^N dx_idy_i dU_1dU_2
$$
 Since $X^2=U_1\diag(x^2_i)U^{-1}_1$, $Y^2=U_2\diag(y^2_i)U^{-1}_2$, then,
using Lemma \ref{L1}, we obtain
\be
\int_{\mathbb{U}_N\times\mathbb{U}_N}\tau_r(N;X^2Y^2,\mathbb{I}_N) dU_1 dU_2=
 K_N\frac{\det\left[\tau_r(N;x^2_iy^2_j,1)  \right]_{i,j}}{\Delta(x^2)\Delta(y^2)}
\ee

Bearing in mind Remark \ref{main_idea} we see that
 the whole integral (\ref{I_U}) is the integral over $2N$ eigenvalues $\{x_i\}$ and $\{y_i\}$:
$$
I^{A^1,A^2,r,v,u}_{\mathbb{U}_N\times\mathbb{U}_N}=\sum_{\alpha,\beta\in\DP}
Q_\alpha(\pb^1_{\rm odd})Q_\beta(\pb^1_{\rm odd})A^1_{\alpha,\gamma}A^1_{\delta,\beta}
<Q_\gamma,Q_{\delta}>^{r,v,u}_{\mathbb{U}_N\times\mathbb{U}_N}
$$
where
$$
<Q_\gamma,Q_{\delta}>^{r,v,u}_{\mathbb{U}_N\times\mathbb{U}_N}=
\int Q_\gamma(X)Q_\delta(Y)\Delta^*(\xb)\Delta^*(\yb)\prod_{i=1}^N \tau_r(1;x^2_iy^2_i,1)\tilde{v}(x_i)
\tilde{u}(y_i)d\Omega_1(x_i,y_i)
=g^{r,\tilde{v},\tilde{u}}_{\gamma,\delta}
$$
where according to Lemma \ref{<QQ>} the entries of the matrix $ g^{r,v,u} $ are
\be\label{moments1}
 g^{r,u,v}_{i,j}=-\frac{1}{2\pi^2}\oint \oint x^{-i}y^{-j} (xy)^{\frac12 N(1-N)}\tau_r(1;x^2y^2)u(x)v(y) dx dy
\ee
The factor $(xy)^{\frac12 N(1-N)}$ appeared thanks to the ratio 
$\frac{|\Delta(\xb)\Delta(\yb)|^2}{\Delta(x^2)\Delta(y^2)}$ where $|x_i|=|y_i|=1$.
We get
\be
I^{A^1,A^2,r,v,u}_{\mathbb{U}_N\times\mathbb{U}_N}(\pb^1_{\rm odd},\pb^2_{\rm odd})
=\sum_{\alpha,\beta\in\DP}Q_\alpha(\pb^1_{\rm odd})Q_\beta(\pb^2_{\rm odd})
\sum_{\gamma,\delta\in\DP}A^1_{\alpha,\gamma}g^{r,\tilde{v},\tilde{u}}_{\gamma,\delta}A^2_{\delta,\beta}
\ee

 \bx \label{ex1} Take $A^1=A^2=\mathbb{I}_\infty$.
 Choose $\tau_r(1;x^2y^2,1)=e^{x^2y^2}$. Then $g_{2i,2j}=\delta_{i,j}\frac{1}{i!}$ and
  $g_{2i-1,j}=0=g_{j,2i-1}$ for each pair $i,j$.
 \ex
 
 \bx \label{ex2} Take $A^1=A^2=\mathbb{I}_\infty$.
 Choose $\tau_r(1;x^2y^2,1)=(1+x^2y^2)^a$. Then $g_{2i,2j}=\delta_{i,j} a(a-1)\cdots(a-i+1)\frac{1}{i!}$ and
  $g_{2i-1,j}=0=g_{j,2i-1}$ for each pair $i,j$.
 \ex
 Actually the moment matrix $g$ is diagonal in these examples where $u(x)v(y)$ is a function
 of the product $xy$. For more general we take different choice, for possible corrections to $v,u$ see also Remark 
 \ref{shift} below.
 
 \bx \label{ZZ2}
 Perhaps, the simplest example is as follows. We take each $A^a$ be identity matrices, $v=u \equiv 1$ 
 and $r(j)=j^{-1}$ which gives
 $$
 \int_{\mathbb{U}_N\times\mathbb{U}_N} 
 e^{\ttr X^2Y^2+\sum_{m>0,{\rm odd}}\frac2m \left(p^{(1)}_m\ttr X^{-m} + p^{(2)}_m\ttr Y^{-m}\right)}d_*Xd_*Y
 =\sum_{\alpha\in\DP}2^{-\ell(\alpha)} Q_{2\alpha}(\pb^1_{\rm odd})Q_{2\alpha}(\pb^2_{\rm odd})
 \prod_{i=1}^{\ell(\alpha)}\frac{1}{\alpha_i!}
 $$
 The right hand side is a special (degenerate) case of the BKP hypergeometric tau function studied in
 \cite{Or}.
 
 \br
It follows from the last formula that in the considered case we get 
(up to an overall factor)
$$
<Q_\alpha(U^\dag_1),Q_\beta(U^\dag_2)>^{}_{\mathbb{U}_N\times\mathbb{U}_N}=
$$
$$
\int_{\mathbb{U}_N\times\mathbb{U}_N} 2^{-\ell(\alpha)-\ell(\beta)}Q_\alpha(U^\dag_1)
\sum_\gamma 2^{-\ell(\gamma)}Q_\gamma(U^2_1 U^2_2)Q_\gamma(\tfrac12\pb_\infty) Q_\beta(U^\dag_2)d_*U_1 d_*U_2
$$
$$
=\begin{cases}
\delta_{\alpha,\beta} 2^{-\ell(\alpha)} \prod_{i=1}^{\ell(\alpha)}\frac{1}{(\frac12 \alpha_i)!},\,
{\rm if\,each}\,\alpha_i\,
{\rm is\, even}\\
0\quad {\rm otherwise} 
\end{cases}
$$
One may ask: is the following formula correct
$$
\int_{\mathbb{U}_N} Q_{2\alpha}(U^\dag)Q_{\beta}(U^2A)d_*U=\delta_{\alpha,\beta}
\frac{Q_\alpha(A)}{Q_\alpha(\pb_\infty)\prod_i {\alpha_i!}} ?
$$
We do not know. 
 \er
 \ex
 
\bx
$$
\int_{\mathbb{U}_N\times\mathbb{U}_N} 
e^{\sum_{m>0,{\rm odd}}\frac2m \left(p^{(1)}_m\ttr X^{-m} + p^{(2)}_m\ttr Y^{-m}\right)}
    \det(1-cX^2_1Y^2)^{-a}   d_*Xd_*Y
$$
$$ 
 =\sum_{\alpha\in\DP} Q_{2\alpha}(\pb^1_{\rm odd})Q_{2\alpha}(\pb^2_{\rm odd})
 \prod_{i=1}^{\ell(\alpha)}\frac{1}{\alpha_i!}
 $$

\ex

\subsection{Both matrices are Hermitian}

\paragraph{A. 2KP case}. The interesting 2KP cases were written in various papers 
\cite{GMMMO},\cite{O-2004-New},\cite{Rangloom}.

In the same way anti-Hermitian matrices are treated.

This case can be considered in the same way as the prevous one. 
It is natural to take the following measure on the space $\mathfrak{H}_N$ of Hermitian $N\times N$ matrices:
$$
 d\Omega_N(X)=C_Ne^{-w_1\ttr X^2}\prod_{i\le j\le N}d\Re X_{i,j} \prod_{i < j\le N}d\Im X_{i,j}  
$$
$$
 d\Omega_N(Y)=C_Ne^{-w_2\ttr Y^2}\prod_{i\le j\le N}d\Re Y_{i,j} \prod_{i < j\le N}d\Im Y_{i,j}  
$$
where $w_i>0$ are parameters.
We take $d\Omega_N(X,Y)=d\Omega_N(X)d\Omega_N(Y)$.
We have $X=U_1\diag(x_i)U^{-1}_1$, $Y=U_2\diag(y_i)U^{-1}_2$,
where $U_1,U_2\in \mathbb{U}_N$. Then we have \cite{Mehta}:
\be
d\Omega_N(X)=c_N (w_1;v)\left(\Delta(\xb) \right)^2 \prod_{i=1}^N e^{-w_1x^2_i}dx_i d_*U
\ee
\be
d\Omega_N(Y)=c_N (w_2; u)\left(\Delta(\yb) \right)^2 \prod_{i=1}^N  e^{-w_2y^2_i}dy_i d_*U
\ee
where we suppose that $\int d\Omega_N(X) = \int d\Omega_N(Y) =1 $.

Repeating the calculation from the previous subsection we obtain
\be\label{J_E}
J^{A^1,A^2,r,v,u,w_1,w_2}_{\mathfrak{H}_N\times\mathfrak{H}_N}(\pb^1,\pb^2):=
\ee
$$
\int_{\mathfrak{H}_N\times\mathfrak{H}_N} \tau_{A^1}(\pb^1,X)\tau_r(N;XY,\mathbb{I}_N)\tau_{A^2}(Y,\pb^2)
\det\left( v(X) u(Y) \right)  d\Omega_N(X) d\Omega_N(Y) 
$$
\be
 =\sum_{\lambda,\mu\in\Pa\atop\ell(\lambda),\ell(\mu)\le N} 
 s_\lambda(\pb^1)s_\mu(\pb^2)\sum_{\tilde{\lambda},\tilde{\mu}\in\Pa}
 A^1_{\lambda,\tilde{\lambda}}(N)g^{r,v,u,w_1,w_2}_{\tilde{\lambda},\tilde{\mu}}(N)A^2_{\tilde{\mu},\mu}(N)
\ee
where $\tau_{A^i},\,i=1,2$ are given by (\ref{tauA1}),(\ref{tauA2}), $\tau_r(N;X Y,\mathbb{I}_N)$ is given by
(\ref{pairing}), $g^{r,v,u}_{\tilde{\lambda},\tilde{\mu}}(N)$ is given by (\ref{g_lambda,mu}) where
according to (\ref{moments-gen}) we have
$$
g^{r,v,u,w_1,w_2}_{i,j}=C\int\int x^{i}y^{j}\tau_r(1;xy,1) v(x)u(y)e^{-w_1x^2-w_2y^2}dx dy
$$
where $\tau_r(1;xy,1)$ is given by (\ref{tau_for_moments}).
The family is parametrized by the choice of $A^i,\,i=1,2$ and by the choice of functions $r,v,u$
and one keeps in mind Remark \ref{gauge_freedom}. Thanks to Lemma \ref{product} the relation 
(\ref{composition_law}) is fulfilled.

Let us present well-known models.
\bx Two-matrix model.
Take $w_1=w_2=0$, $v=u\equiv 1$ and $r(j)=c\sqrt{-1}j^{-1}$ that is
$$
\tau_r(N;XY,\mathbb{I}_N)=e^{c\sqrt{-1}\ttr(XY)},\quad
\tau_r(1;xy,1)=e^{c\sqrt{-1}xy}
$$
Then $
g^{r,v,u,w_1,w_2}_{i,j}=i!c^i 
\delta_{i,j} $
$$
J^{A^1,A^2,r,v,u,w_1,w_2}_{\mathfrak{H}_N\times\mathfrak{H}_N}(\pb^1,\pb^2)=
\sum_{\lambda\in\Pa\atop\ell(\lambda)\le N} s_\lambda(\pb^1)s_\lambda(\pb^2) c^{|\lambda|}
\prod_{(i,j)\in\lambda}\left(N+j-i \right)
$$
This model can be treated as Hermitian-anti-Hermitian two matrix model \cite{HO-Borel} if we put 
$p^{(2)}_m\to (-1)^{\frac12 m} p^{(2)}_m$ and keep $c$ to be real. 

\ex
\bx 
Take $A^1=A^2=\mathbb{I}_\infty$ and  $w_1=w_2=0$, $r(j)=c\sqrt{-1}j^{-1}$ as in the previous example.
We can use the fact $ w_2\ttr Y^2 + c\sqrt{-1}\ttr(XY) - (\sqrt{2}w_2)^{-1} c^2\ttr X^2=
 w_2\sum_{i,j}\tilde{Y}^2$ where $\tilde{Y}$ is the  matrix with entries $\tilde{Y}_{ij}= 
 Y_{ij}+ c\sqrt{-1} X_{ij} $. After the Gaussian integration over  the shifted variables $\{Y_{i,j},\,0\le i,j\le N\}$, we obtain
 \be
 \int_{\frak{H}_N} e^{\sum_{m>0} \frac 1m p^{(1)}_m \ttr X^m + (\sqrt{2}w_2)^{-1} c^2\ttr X^2  }
 d\Omega(X)=\sum_{\lambda\in \Pa\atop \ell(\lambda)\le N} s_\lambda(\pb^1)s_\lambda(0,w_2,0,0,\dots)
 c^{|\lambda|}
\prod_{(i,j)\in\lambda}\left(N+j-i \right)
 \ee
\ex

\paragraph{Unitary-Hermitian two matrix model.}
The same method can obviously be applied to {\it mixed ensemble} of interacting Hermitian and of unitary matrices, 
namely
to the ensemble (\ref{2KP_2KP_2KP=2KP}), where $X\in \mathbb{U}_N$ and $Y\in \mathfrak{H}_N$, for example, for:
\be\label{mixedKP}
\int_{\mathbb{U}_N\times\mathfrak{H}_N} e^{\sum_{m>0}\frac1m \left(p^{(1)}_m\ttr U^{-m} +p^{(2)}_m\ttr X^m \right)
+c \ttr(U X)}\det U^{-k} \det X^n d_*U d\Omega(X)=\sum_{\lambda,\mu\in\Pa\atop\ell(\lambda),\ell(\mu)\le N} 
 s_\lambda(\pb^1)s_\mu(\pb^2)g_{\lambda,\mu}(N)
\ee
where $g_{\lambda,\mu}(N)=\det\left[g_{N+\lambda_i-i,N+\mu_j-j}  \right]_{i,j}$, see (\ref{A_l_m(N)})
and (\ref{moments-gen}) where
$$
g_{i,j}=\frac{(2\pi)^{-\frac32}}{\sqrt{-1}}\oint \frac{dx}{x} \int_{\mathbb{R}} e^{c x y} x^{-k-i}y^{j+n} e^{-\frac12 y^2}dy =
\begin{cases} c^{k+i}\frac{(\frac{i+j+k+n}{2})!!}{(i+k)!} ,\quad i+j+k+n\,{\rm even}\\
 0,\quad i+j+k+n\,{\rm odd}
\end{cases}
$$

The right hand side of (\ref{mixedKP}) has the form of the Takasaki series \cite{Takasaki},\cite{TakashiLMP}
for the TL tau function.

\br
Let us note that there is the direct analogue of the model 
$J^{A^1,A^2,r,v,u,w_1,w_2}_{\mathbb{U}_N\times\mathbb{U}_N}(\pb^1,\pb^2)$
where the unitary matrices $U_1,U^\dag_1$ and $U_2,U^\dag_2$ are replaced by
the complex matrices $Z_1,Z^\dag_1$ and $Z_2,Z^\dag_2$. This model of complex matrices 
together with the model of unitary matrices (\ref{ZZ1}) can be solved 
via character expansion as it was done, say, in \cite{AOV}. In both cases of unitary 
complex and mixed unitary-complex matrix ensembles such models are related to the graph on Figure 1 (c), see  Appendix. 
However, this method 
 does work for unitary matrix models
$I^{A^1,A^2,r,v,u,w_1,w_2}_{\mathbb{U}_N\times\mathbb{U}_N}(\pb^1,\pb^2)$, written
down in (\ref{ZZ2})
and it does 
not directly work for the both Hermitian ensembles 
$J^{A^1,A^2,r,v,u,w_1,w_2}_{\mathfrak{H}_N\times\mathfrak{H}_N}(\pb^1,\pb^2)$ and
$I^{A^1,A^2,r,v,u,w_1,w_2}_{\mathfrak{H}_N\times\mathfrak{H}_N}(\pb^1,\pb^2)$.

\er

\paragraph{B. 2BKP case}

\be\label{I_E}
I^{A^1,A^2,r,v,u,w_1,w_2}_{\mathfrak{H}_N\times\mathfrak{H}_N}(\pb^1,\pb^2):=
\ee
$$
\int_{\mathfrak{H}_N\times\mathfrak{H}_N} \tau^{\rm B}_{A^1}(\pb^1_{\rm odd},X)
\tau_r(N;X^2Y^2,\mathbb{I}_N)\tau^{\rm B}_{A^2}(Y,\pb^2)
\det\left( v(X) u(Y) \right)  d\Omega_N(X) d\Omega_N(Y) 
$$
\be
 =\sum_{\alpha,\beta\in\DP} 
 Q_\alpha(\pb^1_{\rm odd})Q_\beta(\pb^2_{\rm odd})\sum_{\tilde{\gamma},\delta\in\DP}
 A^1_{\alpha,\gamma}g^{r,v,u,w_1,w_2}_{\gamma,\delta}A^2_{\delta,\beta}
\ee
where $\tau_{A^i},\,i=1,2$ are given by (\ref{tauB1}),(\ref{tauB2}), $\tau_r(N;X^2 Y^2,\mathbb{I}_N)$ is given by
(\ref{pairing}), $g^{r,v,u}_{\tilde{\lambda},\tilde{\mu}}(N)$ is given by (\ref{g_lambda,mu}) where
according to (\ref{moments-gen}) we have
$$
g^{r,v,u,w_1,w_2}_{i,j}=C\int\int x^{i}y^{j}\tau_r(1;xy,1) v(x)u(y)e^{-w_1x^2-w_2y^2}dx dy
$$
where $\tau_r(1;xy,1)$ is given by (\ref{tau_for_moments}).
The family is parametrized by the choice of $A^i,\,i=1,2$ and by the choice of functions $r,v,u$
and one keeps in mind Remark \ref{gauge_freedom}. Thanks to Lemma \ref{product} the relation 
(\ref{composition_law}) is fulfilled.

\bx 
Take $A^1,A^2=\mathbb{I}_\infty$ and $w_1=w_2=0$, $v=u\equiv 1$ and $r(j)=c\sqrt{-1}j^{-1}$ that is
$$
\tau_r(N;X^2Y^2,\mathbb{I}_N)=e^{c\sqrt{-1}\ttr(X^2Y^2)},\quad
\tau_r(1;x^2y^2,1)=e^{c\sqrt{-1}x^2y^2}
$$
Then 
\be
g^{r,v,u,w_1,w_2}_{i,j}=\begin{cases}                                           
\left(\frac i2\right)!c^{\frac12 i} \delta_{i,j}, {\rm both}\,i,j\,{\rm even}\\
0\quad {\rm otherwise}
\end{cases}
\ee
We obtain
\be
 \int_{\mathfrak{H}_N\times\mathfrak{H}_N} 
 e^{c\sqrt{-1}\ttr X^2Y^2+\sum_{m>0,{\rm odd}}\frac1m \left(p^{(1)}_m\ttr X^{-m} + p^{(2)}_m\ttr Y^{-m}\right)}
 d\Omega(X)d\Omega(Y)
\ee
 \be
=
\sum_{\alpha\in\DP} 2^{-\ell(\alpha)} Q_{2\alpha}(\pb^1_{\rm odd})Q_{2\alpha}(\pb^2_{\rm odd}) c^{|\alpha|}
\prod_{i=1}^{\ell(\alpha)} \alpha_i!
\ee
\ex

\section{Other ensembles}

\paragraph{Two-matrix model where one matrix is real skew-symmetric and the other is skew-symmetrix Hermitian.}

Let us consider the following two-matrix integral where $X\in\mathfrak{S}_N$ with eigenvalues $\pm x_i\sqrt{-1}$
and $Y$ is skew-symmetric Hermitian (the set of these $N\times N$ matrices we denote $\mathfrak{SH}_N$), 
then its eigenvalues are $\pm y_i$ without $\sqrt{-1}$:
\be
\int_{\mathfrak{S}_N\times\mathfrak{SH}_N} 
e^{\sum_{m>0} \frac1m  \left(p^{(1)}_m\ttr X^{2m} + p^{(2)}_m\ttr Y^{2m} +\ttr \left(XY \right)\right)}
d\omega(X)d\Omega(Y)
\ee
\be\label{2_skew}
=C\sum_{\lambda\in\Pa} s_\lambda(\pb^1)s_\lambda(\pb^2)\prod_{i=1}^{\ell(\lambda)} (2(\lambda_i-i+N))!
\ee
which is the KP tau function of hypergeometric type (\ref{tau-p1-p2}) where $r(i)=(2i)(2i+1)$, therefore
it can be presented in the determinantal form, see Section \ref{Introduction}. The last eqiality (\ref{2_skew})
is derived using (\ref{IZHC_O}) and (\ref{skew_symm_measure}) in Appendix \ref{Ensembles} in 
the similar way as it was done for 
Hermitian-anti-Hermintian two-matrix model in \cite{HO-Borel}.

\paragraph{Two-matrix model where both matrices are skew-symmetric.}
Let us use the following result of Brezin-Hikami \cite{BrezinHikami}
\be
 \int_{\mathbb{O}_{N}}e^{\ttr OXO^{\rm tr}Y}d_*O = 
 \frac{1}{\Delta(\xb^2)\Delta(\yb^2)}\begin{cases}
                                      \det\left[ 2\cosh x_iy_j \right]_{i,j\le m},\,\,N=2m \\
                                      \det\left[ 2\sinh x_iy_j \right]_{i,j\le m},\,\,N=2m+1 
                                     \end{cases}
\ee
where 
\be
d_*O=
\ee
Some other ensembles can be studied in the same way. It will be written down in a more detailed version.
 
 Then we get the following 2KP tau function
\be
I_{\mathfrak{A}_N}(\pb^1,\pb^2):=\int_{\mathfrak{A}_N\times\mathfrak{A}_N}\tau_{A^1}(\pb^1,X)
e^{c\ttr XY}\tau_{A^2}(Y,\pb^2)d\omega(X)d\omega(Y)
\ee
\be
=\sum_{\lambda,\mu} s_\lambda(\pb^1)s_\mu(\pb^1) A^3_{\lambda,\mu}
\ee
where
\be
d\omega
\ee
\be
A^3=A^1 g A^2
\ee
and 
\be
g_{ij}=
\ee

\paragraph{Three matrix models}
One can add the auxilliary complex matrix $Z\in\mathbb{GL}_N(C)$ to the ensemble of the matrices $X$ and $Y$
and consider the three matrix model as follows.
Firstly ae remind the measure on the space of complex $N\times N$ matrices:
\be\label{complex}
d\Omega_N(Z,Z^\dag)=C_N e^{-\ttr ZZ^\dag}\prod_{}d\Re Z_{i,j}d\Im Z_{i,j},
\ee
where $C_N$ is chosen by $\int d\Omega_N(Z,Z^\dag)=1$.
Let us take the same choice of $\tau_{A^i}$, $i=1,2,3$  and
of functions $r,v,u$ as in (\ref{2KP_2KP_2KP=2KP}). And consider 
\be\label{pairingZ}
\tau_r(N;ZXZ^\dag Y,\pb_\infty)=\sum_{\lambda\in\Pa} s_\lambda(ZXZ^\dag Y)s_\lambda(\pb_\infty)
\prod_{(i,j)\in\lambda} r(N+j-i)
\ee
where $\pb_\infty=(1,0,0,\dots)$.

We have 
\be\label{2KP_2KP_2KPZ=2KP}
\int \tau_{A^1}(\pb^1,X^\dag)\tau_r(N;ZXZ^\dag Y,\pb_\infty)\tau_{A^2}(Y^\dag,\pb^2)\det\left(v(X)u(Y) \right) 
d\Omega_N(X,Y)d\Omega(Z,Z^\dag)
\ee
\be
=\tau_{A^3}(\pb^1,\pb^2)
\ee

However let us notice that the auxilliary matrix $Z$ does not add coupling constants to the model
and the right hand side is the same as in (\ref{2KP_2KP_2KP=2KP}) and in this sense it is equivalent 
model.

\paragraph{One-matrix model and 1BKP} We get
$$
\int_{\mathfrak{H}_N\times \mathbb{GL}_N(C)} e^{\sum_{m>0,{\rm odd}}
\frac 1m \left(p^{(1)}\ttr X^m + p^{(2)}\ttr X^{-m}\right) }  \tau_r(N;X^2Y^2,
\pb_\infty) d\Omega(X) =
$$
$$
\frac{1}{\Delta(\yb^2)}
\int_{\mathbb{R}^N}\left( \Delta^*(\xb)\right)^2 
\prod_{i=1}^N e^{\sum_{m>0,{\rm odd}}
\frac 1m \left(p^{(1)}x^m_i + p^{(2)} x^{-m}_i\right) }\tau_r(1;x^2_iy^2_i,1)e^{-wx^2_i}dx_i
$$
which $I_2$ of \cite{HLO}. In particular
$$
\int_{\mathfrak{H}_N\times \mathbb{GL}_N(C)} e^{\sum_{m>0,{\rm odd}}
\frac 1m \left(p^{(1)}\ttr X^m + p^{(2)}\ttr X^{-m}\right) }  \tau_r(N;XZXZ^\dag,
\pb_\infty) d\Omega(X)d\Omega(Z,Z^\dag)=
$$
$$
\int_{\mathbb{R}^N}\left( \Delta^*(\xb)\right)^2 
\prod_{i=1}^N e^{\sum_{m>0,{\rm odd}}
\frac 1m \left(p^{(1)}x^m_i + p^{(2)} x^{-m}_i\right) }\tau_r(1;x^2_i,1)e^{-wx^2_i}dx_i
$$
which is related to the Ising model, see Appendix in \cite{HLO} written by H.Braden.
\paragraph{Three matrix models and 2BKP}
Adding the auxilliary complex matrix $Z\in\mathbb{GL}_N(C)$ as in the 2KP case (\ref{2KP_2KP_2KPZ=2KP})
one obtains 
$$
\int \tau^{\rm B}_{A^1}(\pb^1_{\rm odd},X^\dag)\tau^{\rm B}_r(N;ZX^2Z^\dag Y^2,\pb_\infty)
\tau^{\rm B}_{A^2}(Y^\dag,\pb^2_{\rm odd})
\det\left(v(X)u(Y) \right) 
d\Omega_N(X,Y)d\Omega(Z,Z^\dag)
$$
\be\label{2BKP_2BKP_2BKPZ=2BKP}
=\tau^{\rm B}_{A^3}(\pb^1_{\rm odd},\pb^2_{\rm odd})
\ee
where
\be\label{pairingZ_2}
\tau_r(N;ZX^2Z^\dag Y^2,\pb_\infty)=\sum_{\lambda\in\Pa} s_\lambda(ZX^2Z^\dag Y^2)s_\lambda(\pb_\infty)
\prod_{(i,j)\in\lambda} r(N+j-i)
\ee
and 
where the functions $r,v,u$ and $\tau^{\rm B}_{A^i},\,i=1,2,3$ are the same as in (\ref{2BKP_2KP_2BKP=2BKP}).

\paragraph{Multimatrix models}

As it was mentioned one can use multimatrix models considered in \cite{AOV} to present 
the terms $\tau_r(N;X^2y^2,\mathbb{I}_\infty)$ (or, the term )
describing the pairing of matrices $X$ and $Y$.

Actually any embedded graph drawn on the complex plane can be used provided that what we called 
``the spectrum of the star'' is nontrivial for a single ``star'' results in certain tau function.
By the term "star" we called the top of the graph, blown up to a small circle. Each such vertex has a monodromy: 
the product of the source matrices when walking around the boundary of the star. For details, see the article 
\cite{AOV}, also \cite{NO2020tmp}. One can construct infinitely many graphs, each being  related to a solvable
matrix integral, we present few examples.

It is possible to construct infinitely many graphs equipped with source matrices, each of which specifies a precisely solvable
matrix integral, we give several examples of such models. 

\bx
The $n$-gon gives two monodromies $ U_1\cdots U_n X^2$ and  $Y^2 U^\dag_n\cdots U^\dag_1$ when passing around
inside and outside, and $X^2Y^2$ is the monodromy around a chosen vertex:
\be
\tau_r(N;X^2Y^2,\mathbb{I}_N)=\int_{\mathbb{U}_N\times \cdots \times \mathbb{U}_N} e^{\ttr \left( U_1\cdots U_n X^2+
Y^2 U^\dag_n\cdots U^\dag_1\right) } \prod_{i=1}^n d_*U_i 
\ee
\be
=\sum_{\lambda} s_\lambda(X^2Y^2) s_\lambda(\mathbb{I}_N) \left((N)_\lambda \right)^{-n-1}
\ee
\be
K_N\frac{\det\left[\tau_r(1;x^2_iy^2_j,1)\right]_{i,j}}{\Delta(x^2)\Delta(y^2)}
\ee

Thus we obtain
\be
\int_{\mathbb{U}_N\times \cdots \times \mathbb{U}_N} e^{\sum_{m>0,{\rm odd}}\left(p^{(1)}_m \ttr U^{-m}_{n+1} + p^{(2)}_m \ttr U^{-m}_{n+2} \right)}
 e^{\ttr \left( U_1\cdots U_n U^2_{n+1}+
U^2_{n+2} U^\dag_n\cdots U^\dag_1\right) }\prod_{i=1}^{n+2} d_*U_i
\ee
\be
=K_N\sum_{\alpha\in\DP} 2^{-\ell(\alpha)}Q_\alpha(\pb^1_{\rm odd})Q_\alpha(\pb^2_{\rm odd})
\prod_{i=1}^{\ell(\alpha)}\left(\frac{1}{\alpha_i!}\right)^{n+3}
\ee

\ex

\bx 

\be
\tau_r(N;X^2Y^2,\mathbb{I}_N)=
\int_{\mathbb{GL}_N(C)\times \cdots \times \mathbb{GL}_N(C) } 
e^{\ttr\left( Z_1A_1+\cdots +Z_n A_n  \right)}e^{\ttr\left(Z^\dag_1B_1\right)\cdots \left(Z^\dag_nB_n)  \right)}
\prod_{i=1}^n d\Omega(Z_i,Z^\dag_i)
\ee
where $A_1,\dots,A_n,B_1,\dots,B_n$ are any matrices independent of $\{Z_i,Z^dag_i,i=1,\dots,n\}$ conditioned by
\be
A_1B_1\cdots A_nB_n=X^2Y^2
\ee

\ex

\bx 

\be
\tau_r(N;X^2Y^2,\mathbb{I}_N)=
\int_{\mathbb{GL}_N(C)\times \cdots \times \mathbb{GL}_N(C) } 
e^{\ttr \left(Z_1A_1Z^\dag_1 B_1\right)\cdots \left(Z_nA_nZ^\dag_n B_n\right)}
\prod_{i=1}^n d\Omega(Z_i,Z^\dag_i)
\ee

\ex

\bigskip
\bigskip
\noindent
\small{ {\it Acknowledgements.}
The author is grateful A. Alexandrov, A.Morozov, A. Mironov for
attracting attention to \cite{Alex2},\cite{MMq}, 
P.Gavrilenko for the useful question,
J. Harnad for remarks 
and special thanks to Andrey Mironov for fruitful discussions.
 The work was supported by the Russian Science
Foundation (Grant No.20-12-00195).

\bigskip



\appendix

\section{Partitions. The Schur polynomials}
 
We recall that a nonincreasing set of nonnegative integers $\lambda_1\ge\cdots \ge \lambda_{k}\ge 0$,
we call partition $\lambda=(\lambda_1,\dots,\lambda_{l})$, and $\lambda_i$ are called parts of $\lambda$.
The sum of parts is called the weight $|\lambda|$ of $\lambda$. The number of nonzero parts of $\lambda$
is called the length of $\lambda$, it will be denoted $\ell(\lambda)$. See \cite{Mac} for details.
Partitions will be denoted by Greek letters: $\lambda,\mu,\dots$. The set of all partitions is denoted by
$\Pa$. The set of all partitions with odd parts is denoted $\OP$.
Partitions with distinct parts are called strict partitions, we prefer
letters $\alpha,\beta$ to denote them. The set of all strict partitions will be denoted by $\DP$.
The Frobenius coordinated $\alpha,\beta$ for partitions $(\alpha|\beta)=\lambda\in\Pa$ are of usenames
(let me recall that the coordinates $\alpha=(\alpha_1,\dots,\alpha_k)\in\DP$ consists of the lengths of arms counted
from the main diagonal of the Young diagram of $\lambda$ while
$\beta=(\beta_1,\dots,\beta_k)\in\DP$ consists of the lengths of legs counted
from the main diagonal of the Young diagram of $\lambda$, $k$ is the length of the main diagonal of $\lambda$,
see \cite{Mac} for details).

To define the Schur function $s_\lambda$, $\lambda\in\Pa$ at the first step we introduce the 
set of elementary Schur functions $s_{(m)}$ by
$$
e^{\sum_{m>0}\frac 1m p_m x^m}=\sum_{m\ge 0} x^m s_{(m)}(\pb)
$$
where the variables $\pb=(p_1,p_2,p_3,\dots)$ are called power sum variables. For 
$\lambda=(\lambda_1,\lambda_2,\dots)\in\Pa$
we define 
\be
s_\lambda(\pb)=\det\left[s_{(\lambda_i-i+j)}(\pb)  \right]_{i,j>0}
\ee
If we put $p_m=p_m(X)=\ttr X^m$ where $X$ is a matrix we write $s_\lambda(\pb(X))=s_\lambda(X)$.

To define the projective Schur function $Q_\alpha$ where $\alpha\in\DP$, at the first step,
we define the set of functions $\{q_i,i\ge 0\}$ by
\be
e^{\sum_{m>0,{\rm odd}} \frac 2m p_m x^m} =\sum_{m\ge 0} x^m q_m(\pb_{\rm odd})
\ee
where now $\pb_{\rm odd}=(p_1,p_3,\dots)$.
Next we defined the following skew symmetric matrix
\be
Q_{ij}(\pb_{\rm odd}) := 
\begin{cases}
q_i(\pb_{\rm odd}) q_j(\pb_{\rm odd}) + 2\sum_{k=1}^j (-1)^kq_{i+k}(\pb_{\rm odd}) q_{j-k}(\pb_{\rm odd}) \quad \text{if } (i,j) \neq (0,0), \\
0  \quad \text{if } (i,j) = (0,0) ,
\end{cases} 
\ee
In particular
$
Q_{(j,0)}(\pb_{\rm odd}) = -Q_{(0,j)}(\pb_{\rm odd})=q_j(\pb_{\rm odd}) \ \text{ for } j\ge 1.
$
For a strict partition $\alpha=(\alpha_1,\dots,\alpha_{2r})$ where $\alpha_{2r}\ge 0$
the projective Schur function is defined
\be
Q_\alpha(\pb_{\rm odd}):= \Pf\left[Q_{\alpha_i \alpha_j}(\pb_{\rm odd})\right]_{1\le i, j \le 2r},\quad 
Q_{\emptyset} :=1
\label{Q+_pfaff}
\ee
Let $X$ be a matrix.
If we put $p_m=p_m(X)=\ttr\left( X^m - (-X)^m\right)$ which we call odd power sum variables, $m$ odd
we write $Q_\alpha(\pb_{\rm odd}(X))=Q_\alpha(X)$.

\section{Tau functions and free fermions}

The tau functions (\ref{tauA1})-(\ref{tauA3}) of the two-component KP hierarchy which are of use in our construction
have the following fermionic vacuum expectation value 
\be\label{tau_ferm}
\tau_{A^a}(N,\pb^1,\pb^2) =\langle N,-N| 
e^{\sum_{i,j\ge 0} {A^a}_{i,j}\psi^{(1)}_i(\pb^1)\psi^{\dag(2)}_{-j-1}(\pb^2)}|0,0\rangle,\quad=1,2,3
\ee
where $\pb^a=(p^{(a)}_1,p^{(a)}_2,p^{(a)}_3,\dots)$. We have $ p^{(a)}_m=mt^{(a)}_m$ where 
$t^{(a)}_m$ are the higher times of the two-component KP hierarchy.
Tau functions (\ref{tauB1})-(\ref{tauB3}) can be written as
\be\label{tauB_ferm}
\sum_{N>0} \kappa^N \tau^{\rB}_A(N,\pb^1_{\rm odd},\pb^2_{\rm odd}) =\langle 0|
e^{\kappa\sum_{i,j\ge 0} A^a_{i,j}\phi^{(1)}_i(\pb^1_{\rm odd})\phi^{(2)}_{j}(\pb^2_{\rm odd})}|0\rangle,\quad=1,2,3
\ee
which we can call the determinantal 2BKP tau functions and 
where odd indexed power sums $\pb^a=(p^{(a)}_1,p^{(a)}_3,p^{(a)}_5,\dots)$ are related to the BKP higher
times $\{t^{(a)}_m \}$ by $t^{(a)}_m=\tfrac{2}{m}p^{(a)}_m,\,m\,{\rm odd}$.

Such tau function are rather simple examples of tau functions of two-component hierarchies.

Tau function $\tau_r$ and its fermionic representation was defined in \cite{OS2000} and was
called hypergeometric tau function. It can be also expressed as the following two-component
KP tau function
\be
\tau_r(N;\pb^1,\pb^2)=K_N\langle N,-N| 
e^{\sum_{i\i} e^{-T_i}\psi^{(1)}_i(\pb^1)\psi^{\dag(2)}_{-j-1}(\pb^2)}|0,0\rangle
\ee
where $e^{-T_i},\,i\i$ is related to $r(i),\,i\i$ by $r(i)=e^{T_{i-1}-T_i}$.
(Earlier in different form it was introduced in
\cite{GKM}).

Notations are explained below.

\paragraph{Charged and neutral fermions \cite{JM}.}
The creation and annihilation Fermi modes  satisfy the anticommutation  relations:
\be
[\psi^{a}_j,\psi^{b}_k]_+= [\psi^{\dag a}_j,\psi^{\dag b}_k]_+=0,
\quad [\psi^{a}_j,\psi^{\dag b}_k]_+=\delta_{a,b}\delta_{jk} .
\label{charged-canonical}
\ee
where the superscripts $a$ and $b$ take values 1 and 2 for two-component fermions.
The Fermi fields $\psi^a(z)$, $\psi^{\dag a}(z)$ and the Fermi modes are related by 
$\psi^a(z)=\sum_{i\i} z^i\psi^a_i$, $\psi^{\dag a}(z)=\sum_{i\i} z^i\psi^{\dag a}_{-i-1}$.
One can introduce 'time dependent' Fermi fields as $\psi^a(z,\pb^a)=e^{\sum_{m>0} \frac1m z^m p^{(a)}_m}\psi^a(z)$,
$\psi^{\dag a}(z,\pb^a)=e^{-\sum_{m>0} \frac1m z^m p^{(a)}_m}\psi^{\dag a}(z)$
and the related 'time dependent' Fermi modes $\psi^a(z,\pb^a)$ by 
\be
\psi^a(z,\pb^a)=:\sum_{i\i} z^i\psi^a_i(\pb^a),\quad
\psi^{\dag a}(z,\pb^a)=:\sum_{i\i} z^i\psi^{\dag a}_{-i-1}(\pb^a)
\ee

There are left and right vacuum vectors:
\be
\langle 0|\psi^{a}_{-i-1}=\langle 0|\psi^{\dag a}_{i}=0=\psi^{a}_{i}|0\rangle=\psi^{\dag a}_{-i-1}|0\rangle,
\quad i<0
\ee
We need the left vacuum vectors with Dirac sea levels $N$ and $-N$ for, resectively, first and second
components of Fermi modes which we will define as follows:
\be
\langle N,-N|=\langle 0|\prod_{i=1}^{N}\psi^2_{-i} \psi^{\dag 1}_{i-1}
\ee
\be
\langle N,-N|\psi^{1}_{-i-1}=\langle N,-N|\psi^{\dag 1}_{i}=
\langle N,-N|\psi^{2}_{-j-1}=\langle N,-N|\psi^{\dag 2}_{j}=0,
\quad i<N,\, j<-N
\ee

The creation and annihilation modes of neutral Fermi fields  satisfy the anticommutation  relations:
\be
[\phi^{a}_j,\phi^{b}_k]_+=(-1)^i \delta_{a,b}\delta_{i+j,0}
\label{neutral-canonical}
\ee
where the superscripts $a$ and $b$ take values 1 and 2 for two-component neutral fermions.
In particular $\left(\phi^a_0\right)^2=\frac 12 $, $a=1,2$.
The Fermi fields $\phi^a(z)$ and the Fermi modes are related by 
$\phi^a(z)=\sum_{i\i} z^i\phi^a_i$.
There are left and right vacuum vectors:
\be
\langle 0|\phi^{a}_{-i-1}=0=\phi^{a}_{i}|0\rangle,
\quad i<0
\ee
and as one can verify we get
\be
\langle 0|\phi^{a}_{-i-1}(\pb^a)=0=\phi^{a}_{i}(\pb^a)|0\rangle,
\quad i<0
\ee

The neutral Fermi modes $\phi_j$  can be related to the charged ones by
 \be
\phi^{a}_j :=\frac{\psi^{a}_j+ (-1)^j\psi^{\dag a}_{-j}}{\sqrt 2},\quad j \in \Zb
\label{charged-neutral}
\ee
In particular we have $ \phi^{a}_0|0\rangle  =
\tfrac{1}{\sqrt{2}} \psi^{a}_0|0\rangle $ and $  \langle 0| \phi^{a}_0 =
\tfrac{1}{\sqrt{2}} \langle  0 | \psi_0^{\dag a}  $.

Let us use the following notations. For $\lambda=(\lambda_1,\dots,\lambda_k)\in\Pa$, $\lambda_k >0$
\be\label{Psi-lambda}
\Psi_{\lambda,\mu,N} := \prod_{i=1}^k
\psi^1_{\lambda_i-i+N}\psi^{\dag 2}_{\mu_i-i+N}
\ee
where $\lambda=(\alpha|\beta)$. Note that $|\lambda\rangle = \Psi_{\alpha,\beta}|0 \rangle$,
$\langle \lambda|=\langle 0|\Psi^\dag_{\alpha,\beta}$.

For $\alpha=(\alpha_1,\dots,\alpha_k)\in\DP$ introduce
\bea\label{Phi-alpha}
\Phi_\alpha &:=& 2^{\frac k2}\phi_{\alpha_1}\cdots \phi_{\alpha_{k}}\\  
\Phi_{-\alpha} &:=& (-1)^{\sum_{i=1}^k\alpha_i}2^{\frac k2}\phi_{-\alpha_k}\cdots \phi_{-\alpha_{1}}
\eea
\be
\Phi(X)=\phi(x_N)\cdots \phi(x_1),\quad \Phi^*(Y)=\phi(-y_1^{-1})\cdots \phi(-y_N^{-1})
\ee

We have Sato formula:
\be
s_\lambda(\xb)\Delta(\xb)=\langle 0|\psi^\dag(x_1^{-1})\cdots \psi^\dag(x_N^{-1})\Psi_{\alpha,\beta} |N\rangle =
\langle N|\Psi^\dag_{\alpha,\beta} \psi(x_1)\cdots \psi(x_N) |0\rangle 
\ee
and formula obtained in \cite{You}:
\be\label{Q(x)}
Q_\alpha(\xb)\Delta^*(\xb)=\langle 0|\phi(-x_1^{-1})\cdots \phi(-x_N^{-1})\Phi_\alpha |0\rangle =
\langle 0|\Phi_{-\alpha} \phi(x_1)\cdots \phi(x_N) |0\rangle 
\ee

\be\label{><}
\frac{2^N}{N!}\int \prod_{i=1}^N d\Omega_1^{r,v,u}(x_i,y_i) \phi(x_N)\cdots \phi(x_1)|0\rangle
\langle 0|\phi(-y_1^{-1})\cdots \phi(-y_N^{-1}) =
\sum_{\alpha,\beta\in\DP\atop\ell(\alpha),\ell(\beta)\le N}\Phi_\alpha|0\rangle g^{r,v,u}_{\alpha,\beta}
\langle 0|\Phi_{-\beta}
\ee

 \paragraph{Examples of tau functions}

 \paragraph{Vacuum tau functions}

Notice that
$$
e^{\sum_m t_m\ttr X^m}\quad {\rm and}\quad e^{-\ttr XY}
$$
can be viewed as simplest (``vacuum'') Toda lattice tau function !

\paragraph{Sato-Takasaki series}

2KP (-Toda lattice) tau function,
(Sato,Takasaki,Takebe):
$$
\tau_N({\bf t},{\bf t}')=\sum_{\lambda,\mu} s_\lambda({\bf t}) g_{\lambda,\mu}(N) s_\mu({\bf t}')
$$
where 

 \begin{itemize}
  \item sum ranges over all possible pairs of partitions $\lambda,\mu$
  \item $s_\lambda , s_\mu$ are the Schur function (polynomials in KP higher times)
  \item $g_{\lambda,\mu}$ is the determinant of a certain matrix (initial data for the TL solution)  
 \end{itemize}
 
{\bf Example}
$$
e^{\sum_m t_m \ttr X^m}=\sum_{\lambda} s_\lambda({\bf t})s_\lambda(X)
$$
$$
e^{\ttr XY}=\sum_{\lambda} s_\lambda(XY) s_\lambda(1,0,0,\dots)
$$

\section{Matrix ensembles \label{Ensembles}}

\paragraph{Unitary matrices.}
The Haar measure on $\mathbb{U}_N$ in the explicit form is written as
\be
d_*U=\frac{1}{(2\pi)^N}\prod_{ 1\le i<k\le N} |e^{\theta_i}-e^{\theta_k}|^2 \prod_{i=1}^N d\theta_i,\quad 
-\pi < \theta_1 < \cdots <\theta_N \le \pi
\ee

\paragraph{Hermitian matrices.}

This case can be considered in the same way as the prevous one. 
It is natural \cite{Mehta} to take the following measure on the space $\mathfrak{H}_N$ of Hermitian $N\times N$ matrices:
$$
 d\Omega_{N,w}(X)=C_N\prod_{i\le j\le N}e^{-w\left(\Re X_{i,j}\right)^2 }d\Re X_{i,j} 
 \prod_{i < j\le N} e^{-w\left(\Im X_{i,j}\right)^2 } d\Im X_{i,j}  
$$
$$
 = c_N \left(\Delta(\xb)\right)^2  \prod_{i=1}^N e^{-w x^2_i}dx_i d_*U
$$
where $w>0$ is a parameter and where we use 
 $X=U\diag\left(\,\,\,0\quad x_i\atop -x_i\,\,0\right)U^{-1}$, $U\in \mathbb{U}_N$.

\br
In many applied problems in which random matrices are used, it is very convenient to assume 
that the parameter $w$ is proportional to the size of the matrices $N$.
\er
  
 \paragraph{Orthogonal matrices.}
The Haar measure on $\mathbb{O}_N$ is 
\be
d_*O=\begin{cases}
\frac{2^{(n-1)^2}}{\pi^n}\prod_{i<j}^n \left(\cos(\theta_i)-\cos(\theta_j)^2  \right)\prod_{i=1}^{n},\quad N=2n
      \\
\frac{2^{n^2}}{\pi^n}\prod_{i<j}^n \left(\cos(\theta_i)-\cos(\theta_j)^2  \right)\prod_{i=1}^{n}
 \sin^2\frac{\theta_i}{2}  d\theta_i ,\quad N=2n+1
     \end{cases}
\ee 
 where
$d\theta_i ,\quad 0\le \theta_i <\cdots<\theta_n \le \pi$. The prefactors are chosen to provide 
$\int_{\mathbb{O}_N} d_*O=1$.
 
By $\mathfrak{S}_N$ we denote the space of $N\times N$ real skew-symmetric matrices.
Recall that the eigenvalues of real skew fields are purely imaginary;
the eigenvalues occur in pairs $\pm x_i\sqrt{-1},\,i=1,\dots,n$  in the case $N=2n$ while
in the case $N=2n+1$ there is an additional eigenvalue $x_{2n+1}=0$.

\paragraph{Skew-symmetric matrices.}
For any $X\in\mathfrak{S}_N$ there exists such $O\in\mathbb{O}_N$ that
$X=O\diag\left\{\left(\,\,\,0\quad x_i\atop -x_i\,\,0\right)_{i=1,\dots,n}\right\}O^{-1}\in\mathfrak{S}_{2n}$,
where $O\in \mathbb{O}_{2n}$,
$X=O\diag\left\{\left(\,\,\,0\quad x_i\atop -x_i\,\,0\right)_{i=1,\dots,n},0\right\}O^{-1}\in\mathfrak{S}_{2n+1}$,
where $O\in \mathbb{O}_{2n+1}$.

The measure on $\mathfrak{S}_N$ is as follows:
$$
d\omega_N(X)=\pi^{\frac12 (n-n^2)}\prod_{i<j}^n e^{-X_{i,j}^2}dX_{i,j}
$$
\be\label{skew_symm_measure}
=\prod_{i<j\le n  } (x^2_i - x^2_j)^2  \prod_{i=1}^n e^{-x^2_i}dx_i d_*O
\begin{cases}
c_1,\quad N=2n
\\
c_2 x^2_i,\quad N=2n+1
\end{cases}
\ee
where prefactors $c_1=2^{-n}\prod_{i=0}^{n-1} \left( \Gamma(2+i)\Gamma(\tfrac12+i) \right)^{-1}$ and 
$c_2=2^{-n}\prod_{i=0}^{n-1} \left( \Gamma(2+i)\Gamma(\tfrac32+i) \right)^{-1}$ 
provide the normalization $\int d\omega_N(X)=1$, see (17.6.5) in \cite{Mehta}.

It is known \cite{PFEFB} that 
\be\label{IZHC_O}
\int_{\mathbb{O}_N} e^{\ttr \left(OXO^{-1}Y\right)} d_*O =\begin{cases}
c_1 \frac{\det\left[2\cosh(2x_iy_i) \right]_{i,j}}{\Delta(\xb^2)\Delta(\yb^2)},\quad N=2n
                                              \\  
c_2 \frac{\det\left[2\sinh(2x_iy_i) \right]_{i,j}}{\Delta(\xb^2)\Delta(\yb^2)\prod_{i=1}^n x_iy_i },\quad N=2n+1                                          
                                             \end{cases}
\ee
where $X,Y\in\mathfrak{S}_N$ and $\pm x_i\sqrt{-1}$ and $\pm y_i\sqrt{-1}$ are eigenvalues of respectively 
$X$ and $Y$.


\paragraph{Complex matrices.}
The measure on the space of complex matrices is defined as 
\be
d\Omega(Z)=c_N \prod_{a,b=1}^N d\Re Z_{ab}d\Im Z_{ab}\text{e}^{-N|Z_{ab}|^2}
\ee

We will consider integrals over $N\times N$ complex matrices $Z_1,\dots,Z_n$ where the measure is defined as
\be\label{CGEns-measure}
d\Omega(Z_1,\dots,Z_n)= c_N^n
\prod_{i=1}^n\prod_{a,b=1}^N d\Re (Z_i)_{ab}d\Im (Z_i)_{ab}\text{e}^{-N|(Z_i)_{ab}|^2}
\ee
where the integration domain is $\mathbb{C}^{N^2}\times \cdots \times\mathbb{C}^{N^2}$ and where $c_N^n$
is the normalization
constant defined via $\int d \Omega(Z_1,\dots,Z_n)=1$.

The set of $n$ $N\times N$ complex matrices
 with measure (\ref{CGEns-measure}) is called the set of {\it $n$ independent complex
Ginibre ensembles}. Such ensembles have wide applications in physics  and in
information transfer theory \cite{Ak1},\cite{Ak2},\cite{AkStrahov}
\cite{S1},\cite{S2},\cite{Alfano}.

\section{Solvable multi-matrix models, related to embedded graphs drawn on the Riemann sphere \cite{AOV}}

 Here we present some examples of tau functions $\tau_{A^i},\i=1,2$, see (\ref{tauA1}),(\ref{tauA2}),
 which are the building blocks of the family (\ref{2KP_2KP_2KP=2KP}) (and also of the chain family (\ref{chain_A})) 
 and some examples of the pairing tau functions $\tau_r(N;XY,\mathbb{I}_N)$ and $\tau_r(N;X^2Y^2,\mathbb{I}_N)$
which are the building bloc
which are the building blocks respectively

Consider a  connected graph $ \Gamma $ on an orientable connected surface $ \Sigma $ without boundary with Euler
characteristic $ \e $.
We require such properties of the graph: 

(1) its edges do not intersect. For example, the edges of the graph in figure () do not intersect: the fact is 
that the graph is drawn on a torus, and not on a piece of paper. 

(2) if we cut the surface $\Sigma$ along the edges of the graph, then the surface will decompose into disks (more 
precisely: into pieces homeomorphic to disks).

As an example, see Figure 1 which contains all such graphs with 2 edges.

  \setlength{\myStandardFigureWidth}{\linewidth}
\setlength{\subSubFigPenalty}{5mm}
\begin{arrangedFigure}{1}{5}{figure1}{All possible graphs with 2 edges (two matrix models). Graph (b) is dual to (c).
Ex means an Example below.}
\subFig[2 edges, 1 vertex, $\e=0$ (torus) Ex 2, Ex 9]{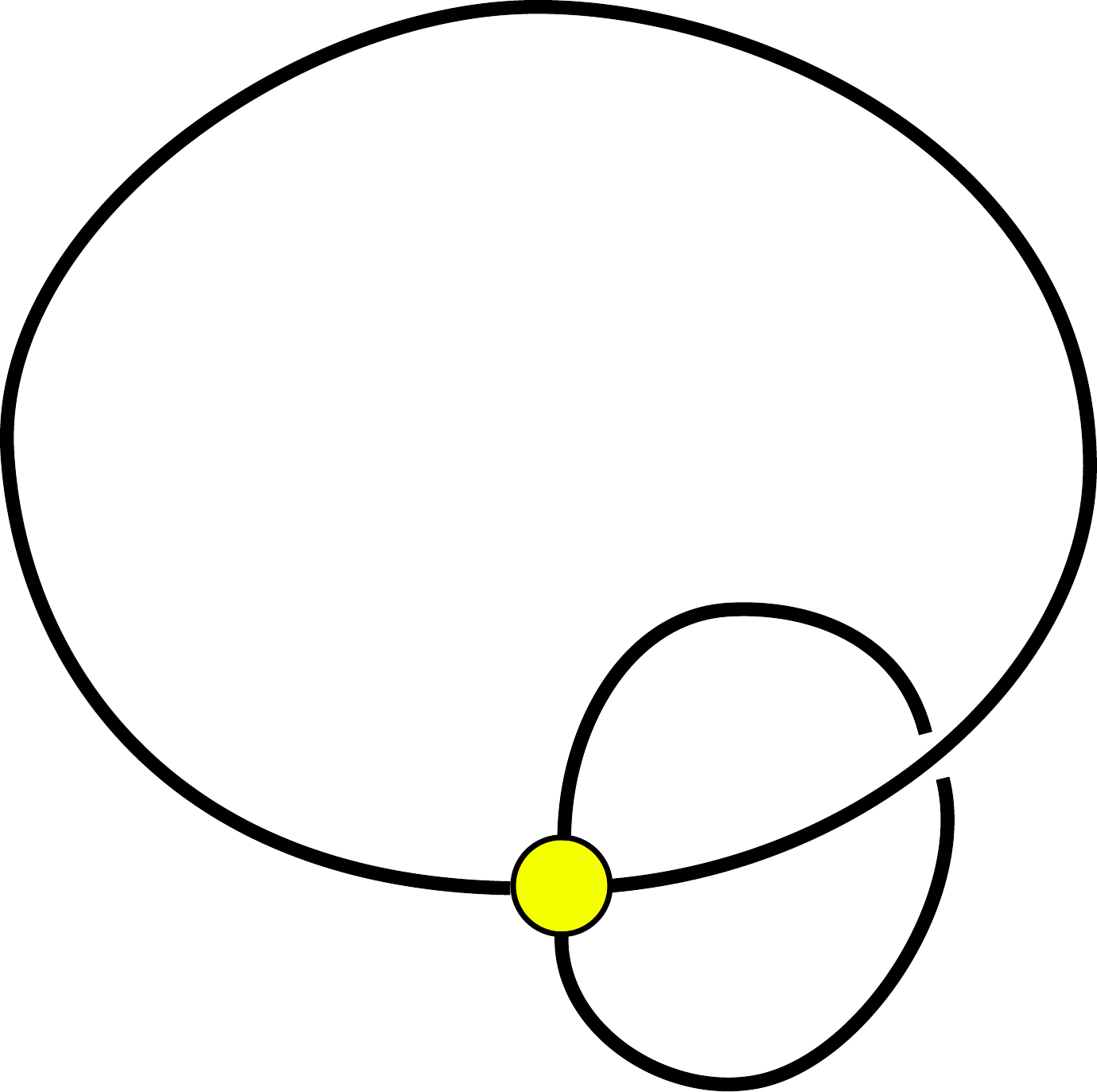}%
\subFig[2 edges, 3 vertices, $\e=2$ (sphere) Ex 6, Ex 8 both]{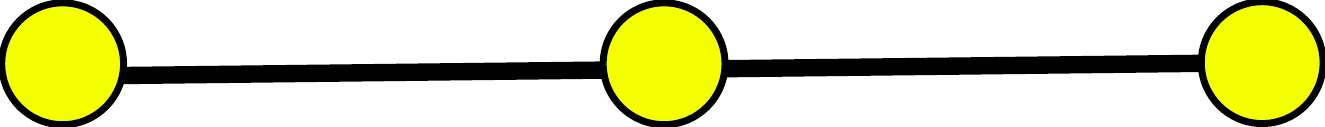}%
\subFig[2 edges, 1 vertex, $\e=2$ (sphere) Ex 6, Ex 8 both]{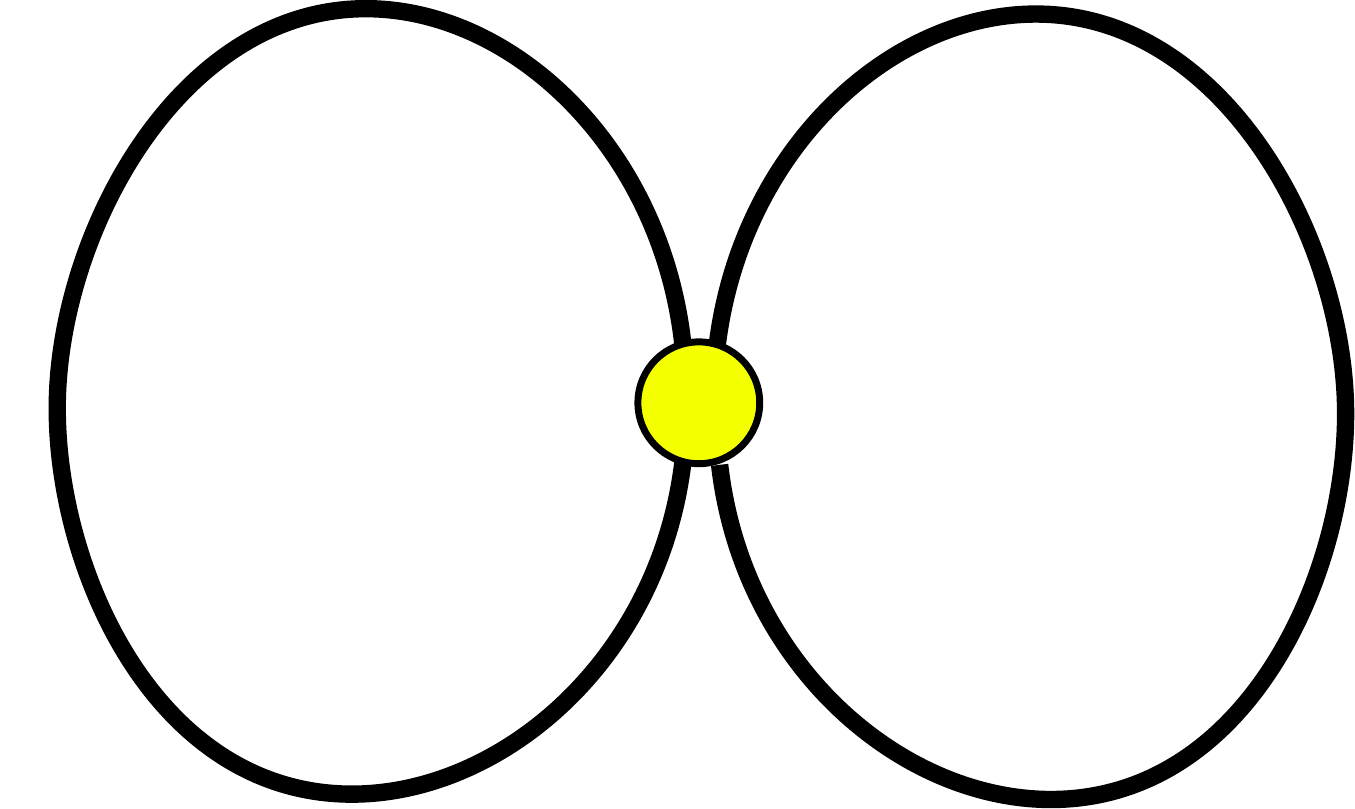}%
\subFig[2 edges, 2 vertices, $\e=2$ (sphere) Ex 7 ]{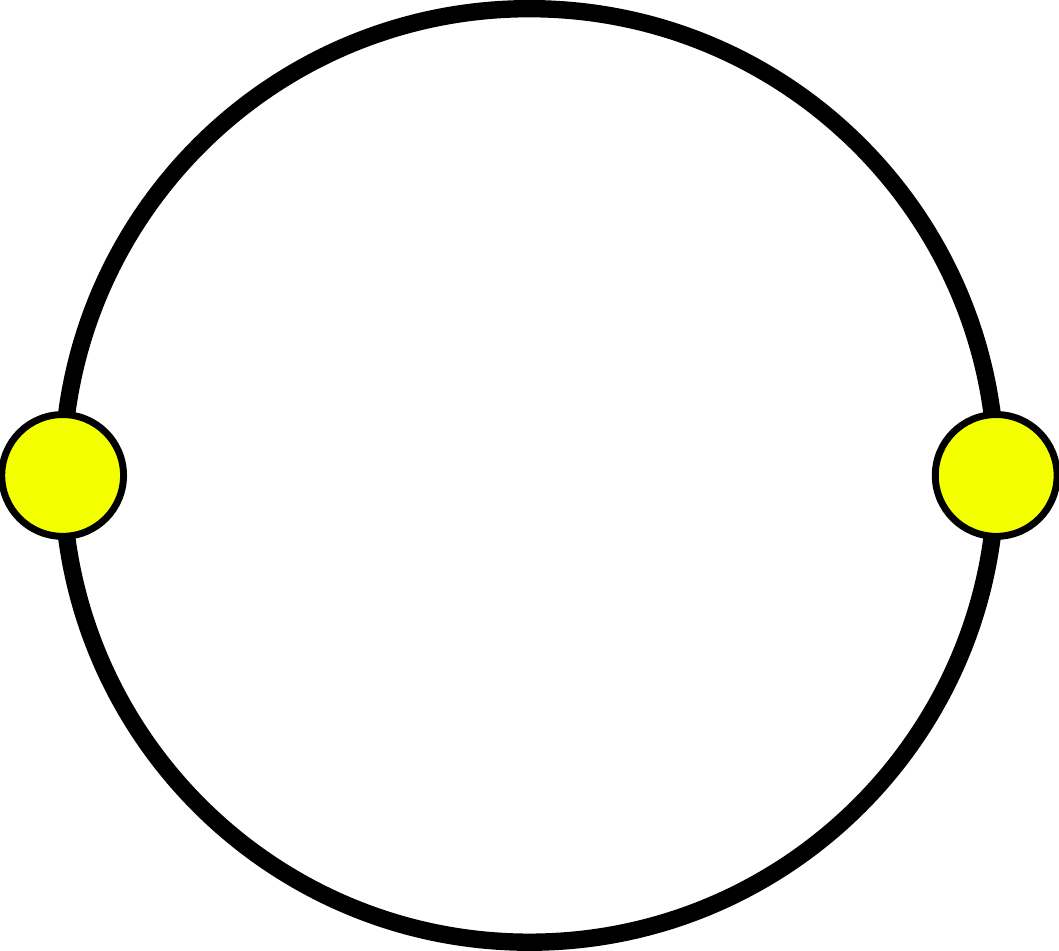}%
\subFig[2 edges, 2 vertices, $\e=2$ (sphere) Ex 4]{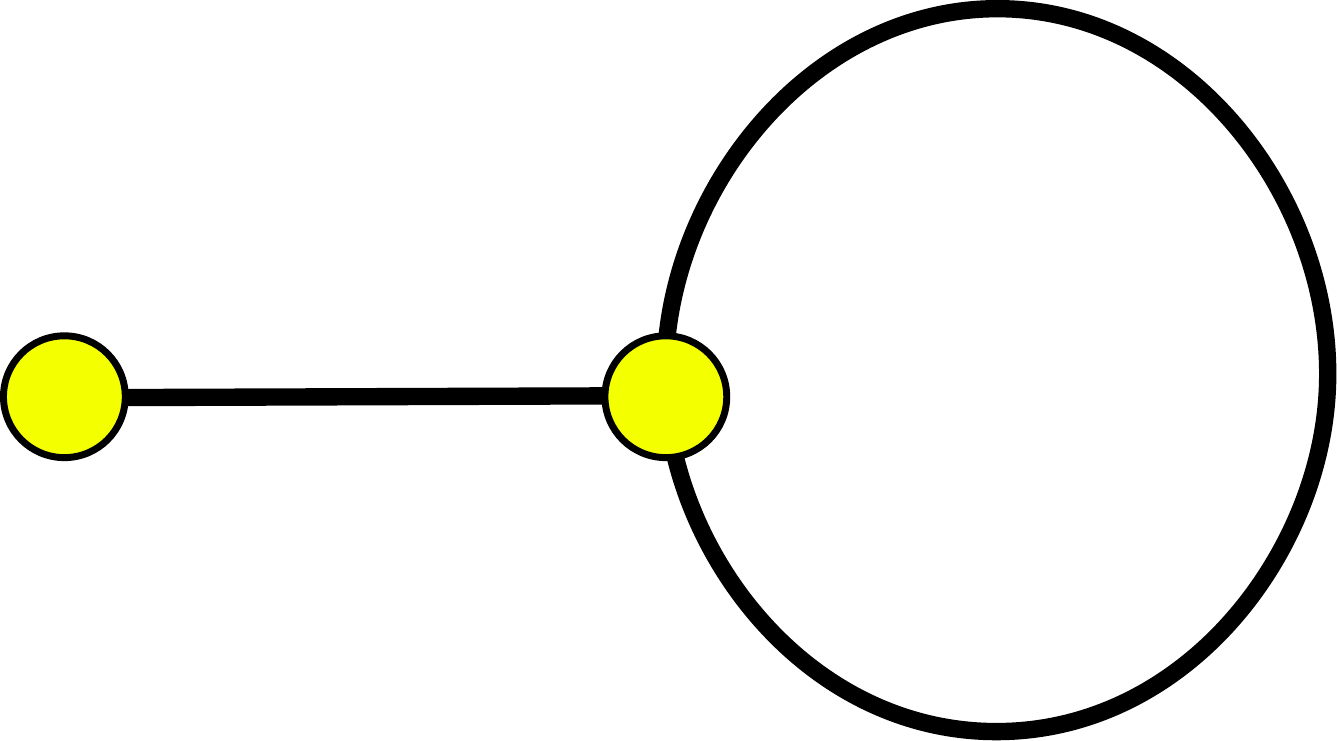}%

\end{arrangedFigure}

Such a graph is sometimes called a (clean) dessins d'enfants\footnote{the term dessins 
d'enfants without the additional "clean" serves for such graph with a bipartite structure}, 
sometimes - a map \cite{LZ}.

Let our graph have $\f$ faces, $n$ edges and $\texttt{v}$ vertices, then $ \e = \texttt{v}- n + \f $.

We number all stars (i.e., all vertices of the graph $\Gamma$) with numbers from $1$ to $\texttt{v}$ and
all faces of $\Gamma$ with numbers from $1$ to $\f$ and all edges of $\Gamma$ with numbers from $1$ to $n$ 
in any way.

We will slightly expand the vertices of the graph and turn them into \textit{small disk}, which sometimes 
for the sake of visual clarity we will call \textit{stars}.

  \setlength{\myStandardFigureWidth}{\linewidth}
\setlength{\subSubFigPenalty}{5mm}
\begin{arrangedFigure}{1}{4}{figure2}{Decorated graphs}
\subFig[1 edges (1-matrix model), 1 vertex (1 star), $\e=0$ (sphere)Ex 1 ]{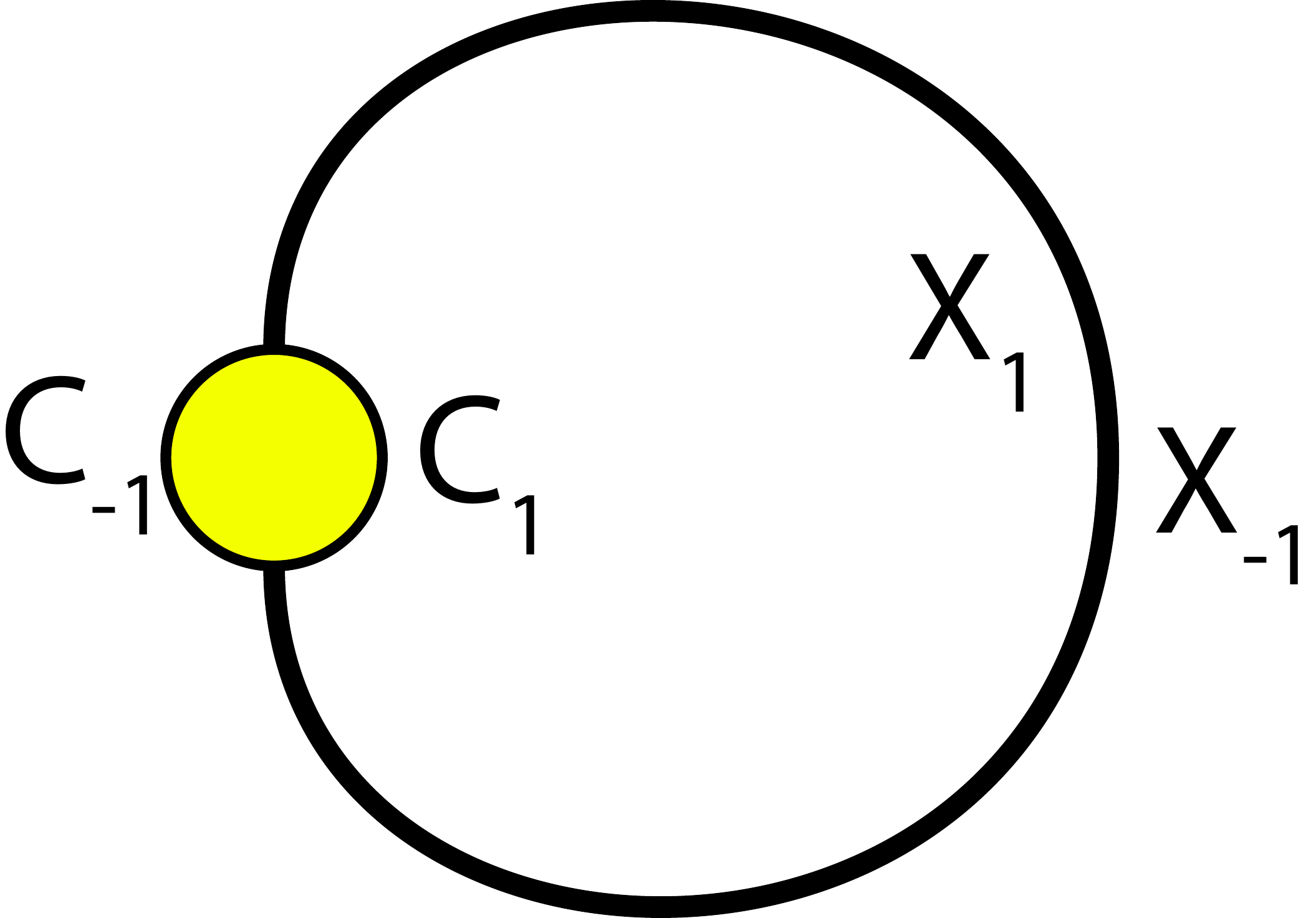}%
\subFig[1 edges, 2 vertices, $\e=2$ (sphere) Ex 1]{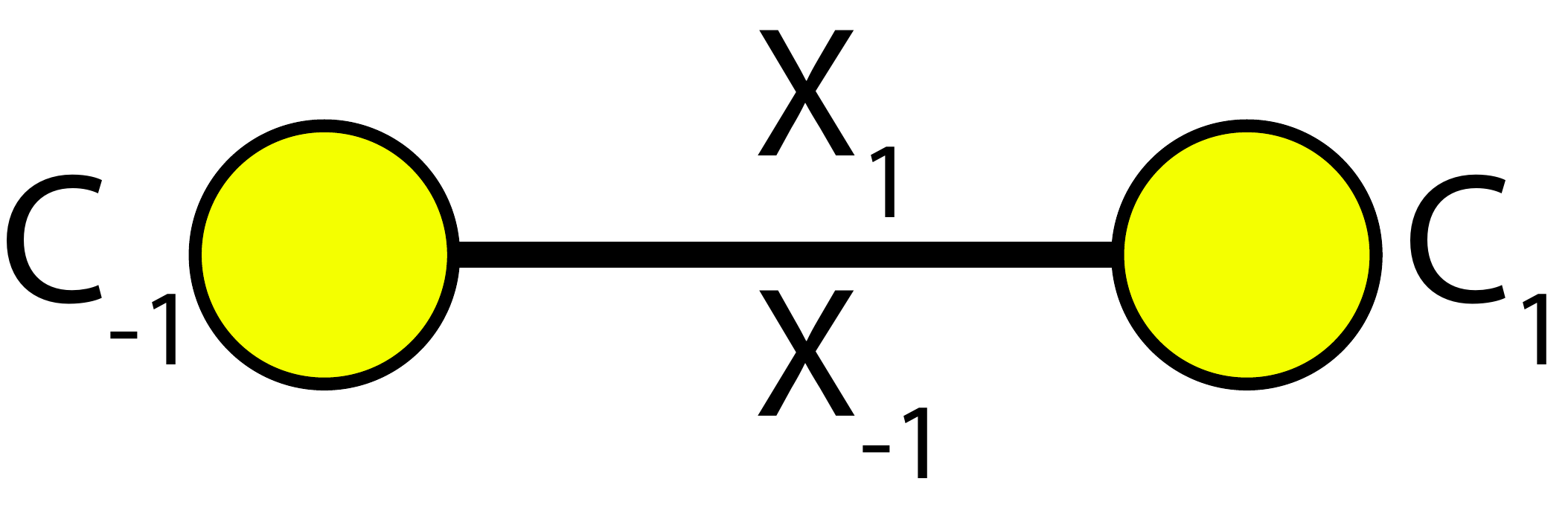}%
\subFig[A fragment of a graph with numbers on the edges ]{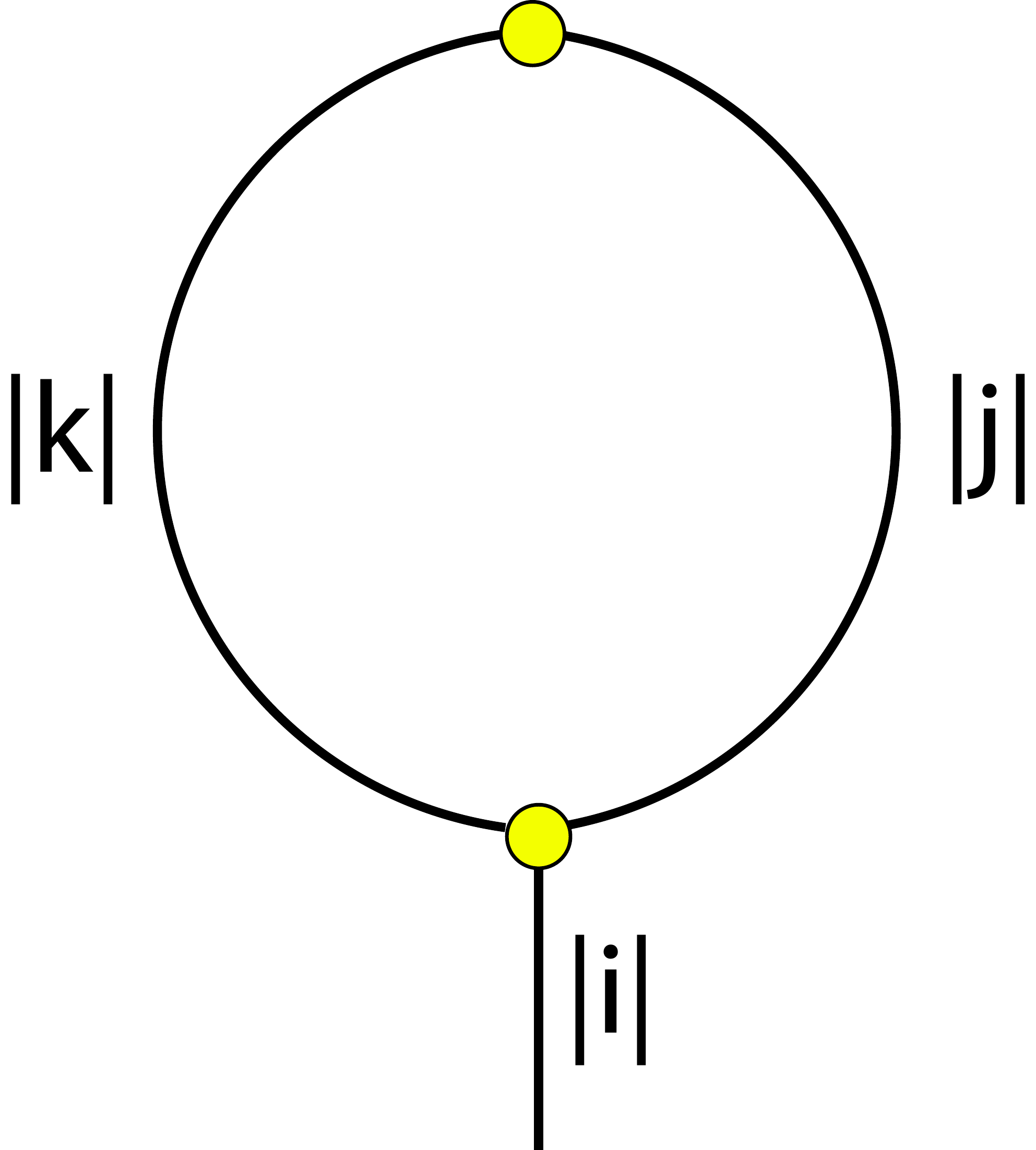}%
\subFig[The same fragment of a graph with random matrices on the edges and source matrices on stars ]{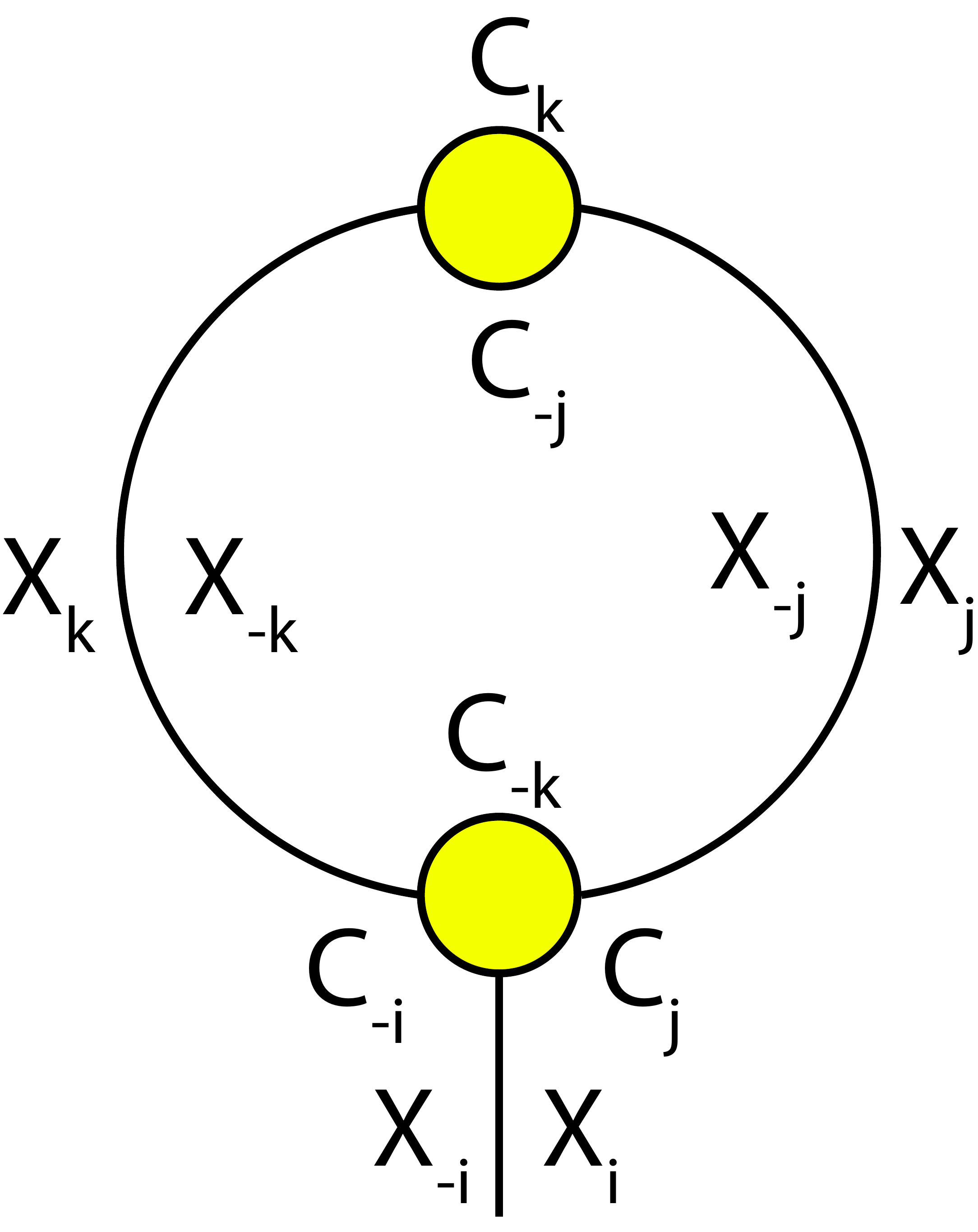}%
\end{arrangedFigure}

In this case, the edges coming out of the vertex will divide
border of a small disk into segments. We orient the boundary of each such segment is clockwise, and the 
segment can be represented 
by an arrow that goes from one edge to another, see the Figures 2 (a), (b), (d) as examples.
Our graph will have a total 
of $2n$ arrow segments.

It is more correct to assume that the edges of the graph are very thin ribbons, that is, they have finite 
thickness.  
That is, the edge number $|i|$ has two sides, one of which we number with the number $i$, and the other with the 
number $-i$ 
(this choice is arbitrary but fixed).
It would be more correct to depict each  edge of  $\Gamma$ numbered by $|i|$
 in the form of a ribbon, the sides of which are two oppositely directed arrows, one arrow has a number $i$, 
 and the second $-i$.

Now, each arrow of the side of the edge rests with its end against the arrow-segment - 
it is as if continued by the arrow-segment. 
We assign the same number $i$ ($i=\pm 1,\dots,\pm n$) to the side of the edge $ \Gamma $ and to the segment of the small disk, which
continues this side while traversing the face in the positive direction.
  
And to the number $i$ ($i=\pm 1,\dots,\pm n$) we attribute the product of the $N\times N$ matrices
$$
i\,\to\,X_iC_i
$$
where $ C_i $ is assigned to the $ i $ segment of a small disk and will be called \textit{source matrix},
and $ X_i $ is assigned to the arrow $ i $, which is the side of the edge $ | i | $ of the graph $ \Gamma $,
see Figure 2 for an illustration.

Let's use the following numbering. If one side of the edge of the tape $ | i | $ is labeled $ i $ then the other 
side of
the same edges is labeled $ -i $. (This doesn't matter which side of the edge $|i|$ we assign the number $i$ to, and 
which - the number $-i$, but we should fix the numbering we have chosen).
The two sides of the edge of the graph $ \Gamma $ are actually arrows looking in opposite directions 
(In order not to complicate the picture, we do not depict these arrows in our figures). 
We attribute
the number $ -i $.
We assign the number $-i$ to both the edge side and the arrow-segment, in which the side $-i$ rests.
Acting in this way, we give numbers to all segments of all small disks.

We attribute a sequential set of numbers to each star as follows.
(this set is defined up to a cyclic permutation, and we will call it a cycle 
associated with the star).
Examples: These are numbers $(1,-1) $ 
assigned to the star (yellow small disk) in  Figure 2 (a). These are
the numbers $(k,-j)$ that we attribute to the star on the top in the Figure 2 (c) and the numbers $(-k,j,-i)$
that we attribute to the lower star in the same figure.
 And to each cycle we ascribe the related cyclic products:
$$
(1,-1)\,\leftrightarrow\,C_1C_{-1},\quad 
$$
to the star (yellow small disk) in Figure 2 (a).  As for the stars in Figure 2 (c) we obtain
$$
(k,-j)\,  \leftrightarrow \,C_kC_{-j}\,\quad (-k,j,-i)\,  \leftrightarrow \,C_{-k}C_{j}C_{-i}
$$
Each cycle product we will call the star monodromy. 
As a result, a star's monodromy is a product of matrices that are attributed to 
arrow segments, taken in the same sequence in which arrows follow each other when moving around a small disk clockwise.
We number the stars in any way by numbers from $1$ to $\texttt{v}$.
The
monodromy of a star $i$ will be denoted by the letter $W_i^*$.

In addition to the edges and in addition to the vertices, we number the 
faces of the graph $\Gamma$ and the corresponding face  monodromies with numbers from $1$ to $\f$.
The cycles corresponding to the face $i$ will be denoted by $f_i$ and defined as follows.
When going around the face border in the positive direction, namely counterclockwise 
for an ordinary face (or counterclockwise if the face contains infinity), we
collect segment numbers of small disks; for a face with a number $i$, this ordered collection of ordered 
numbers is $f_i$. 
As in the case of stars, we build cyclic products  $f_i\leftrightarrow W_i$. Examples:
$$
f_1=(1)\,\leftrightarrow\,C_1=W_1 \quad{\rm and}\quad f_2=(-1)\,\leftrightarrow\,C_{-1}=W_2 
$$
for two faces in Figure 2 (a).

We also introduce dressed monodromies:
$$
{\cal L}_X(W_1)=X_1C_1,\quad {\cal L}_X(W_{-1})=X_{-1}C_{=1}
$$
And for the face in Figure 2 (c):
$$
f_1= (-j,-k)  \, \leftrightarrow \,C_{-j}C_{-k}=W_1\quad \leftrightarrow \Dr_X(W_1)=
X_{-j}C_{-j}X_{-k}C_{-k}
$$

So, we have two sets of cycles and two sets of monodromies: vertex cycles and vertex monodromies
\be
\sigma_i^{-1}\,\leftrightarrow \, W_i^*,\quad i=1,\dots,\texttt{v}
\ee
and face cycles and face monodromies:
The cycles corresponding to the face $i$ will be denoted by $f_i$, we have
\be\label{face-monodromy}
f_i\,\leftrightarrow \, W_i \,\leftrightarrow \, \Dr_X[W_i],\quad i=1,\dots,\f
\ee

Let us note that both cycles and monodromies are defined up the the cyclic permutation.

\br
Important remark. 
Please note that each of the matrices $C_i,\,i=\pm 1,\dots,\pm n$ enters the set of monodromies 
$W_1,\dots,W_f$ once and only once. Accordingly, each random matrix $X_i,\,i=\pm 1,\dots,\pm n$ is included once 
and only once in the set of dressed monodromies $\Dr_X[W_1],\dots,\Dr_X[W_f]$. This determines the class of matrix 
models  that we will consider
and which we will call matrix models of dessins d'enfants.

\er

  \setlength{\myStandardFigureWidth}{\linewidth}
\setlength{\subSubFigPenalty}{5mm}
\begin{arrangedFigure}{1}{4}{figure3}{Graphs drawn without decoration. Graph (a) is dual to (b). Graph (c) is dual to (d)}
\subFig[3 matrix model, $\e=2$ (sphere), Ex 6]{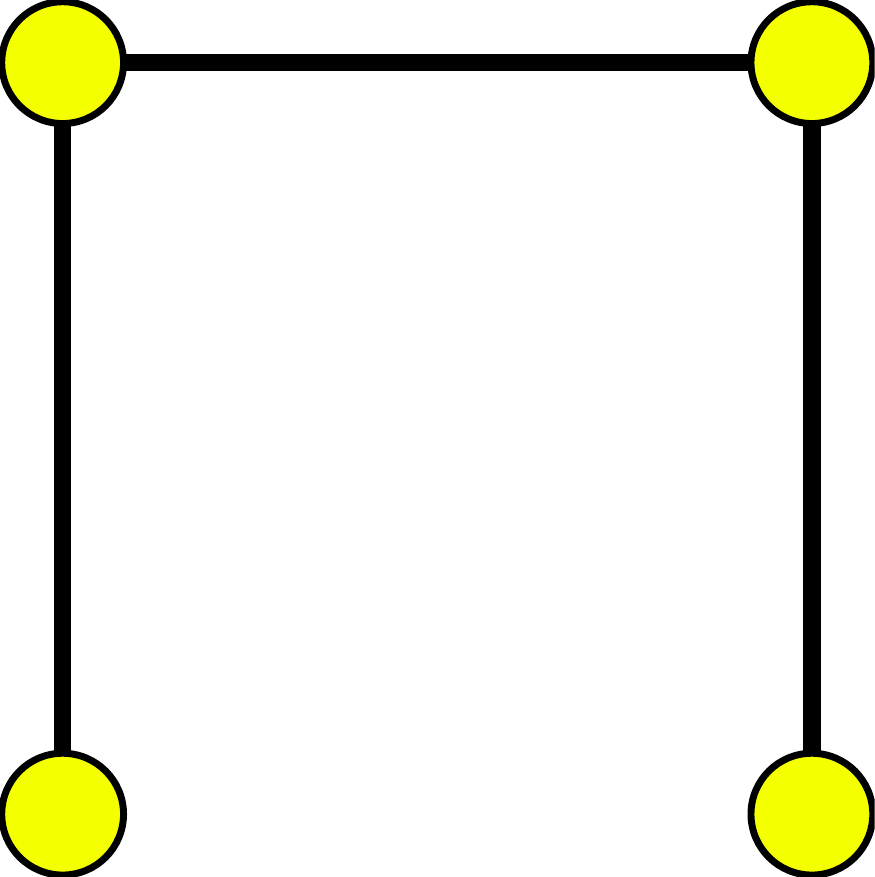}%
\subFig[3 matrix model, $\e=2$ (sphere) Ex 6]{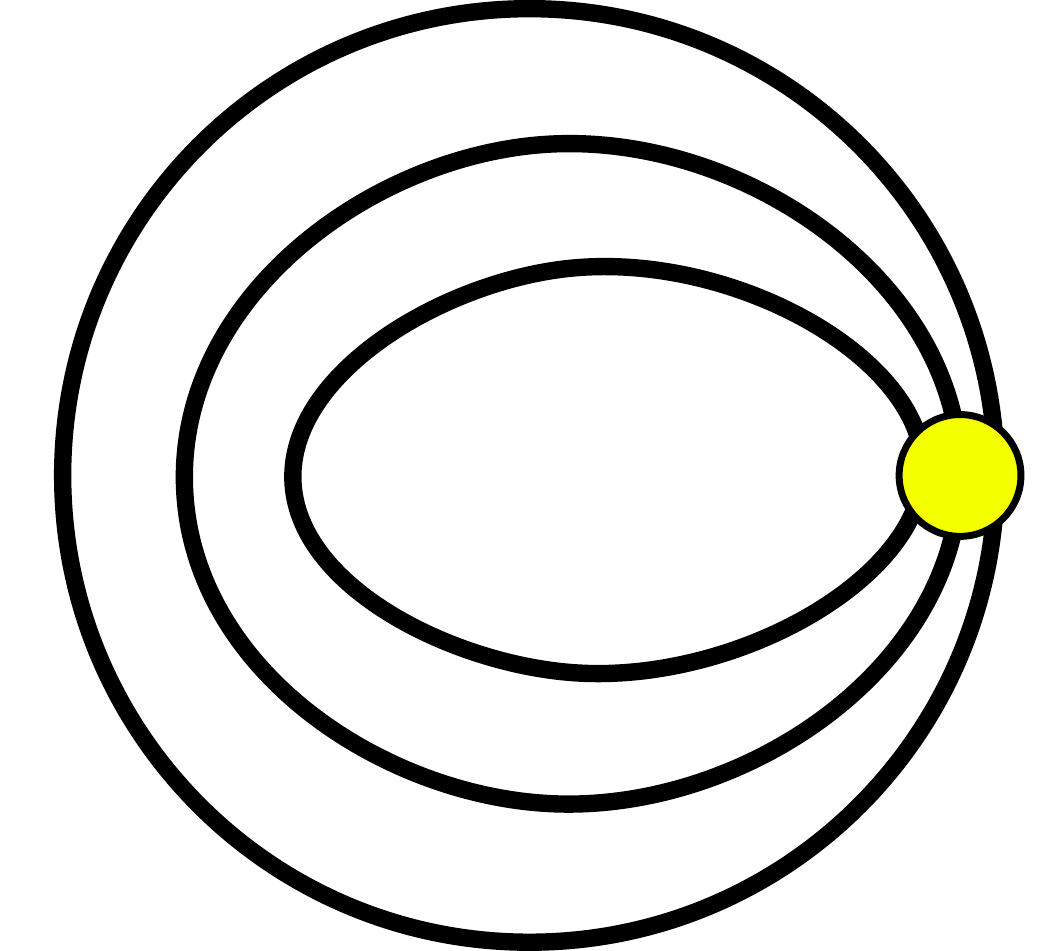}%
\subFig[4 matrix model, $\e=2$ (sphere) Ex 7]{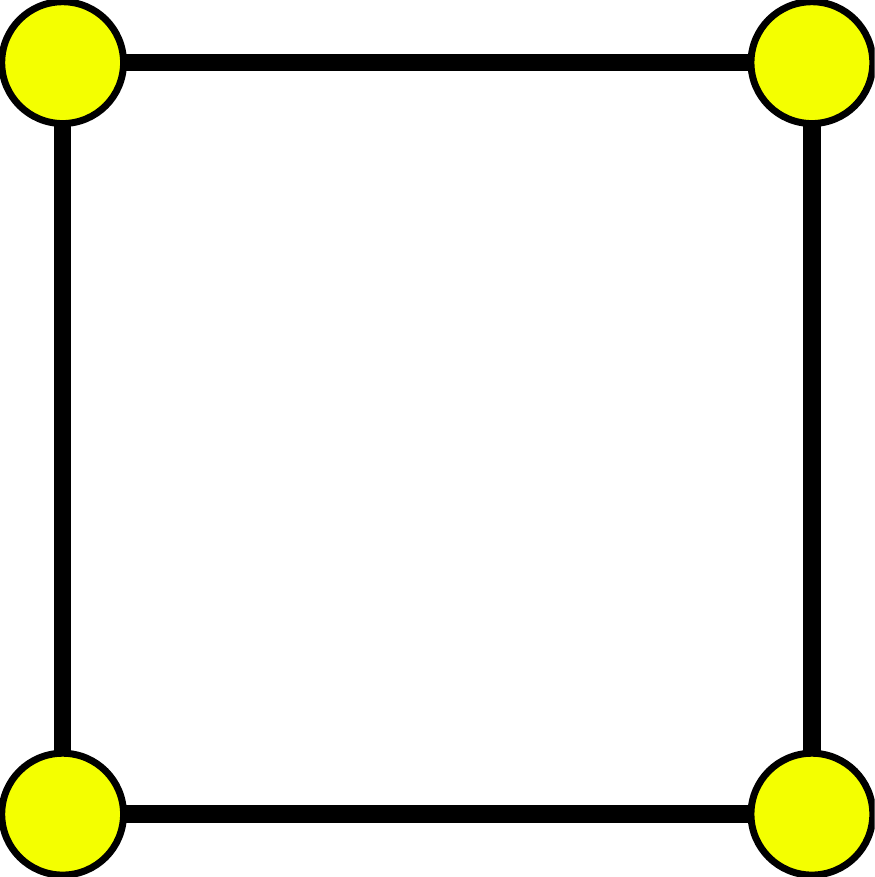}%
\subFig[4 matrix model, $\e=2$ (sphere) Ex 7]{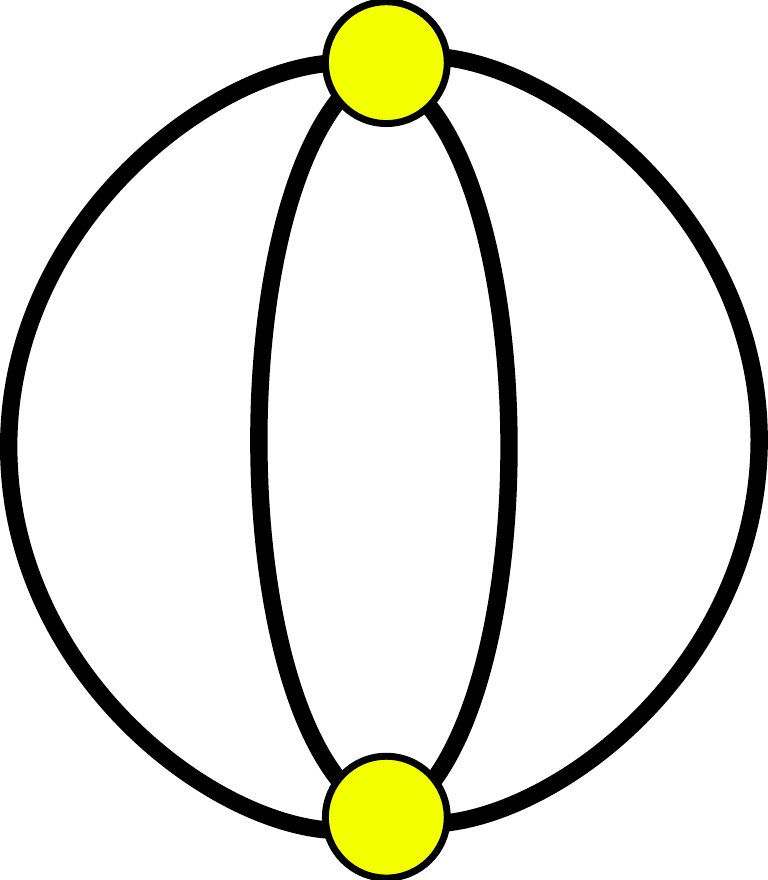}%
\end{arrangedFigure}

  \setlength{\myStandardFigureWidth}{\linewidth}
\setlength{\subSubFigPenalty}{5mm}
\begin{arrangedFigure}{1}{5}{figure4}{Graphs drawn without decoration. Graph (a) is dual to (b)}
\subFig[3 matrix model, $\e=2$ (sphere) Ex 8]{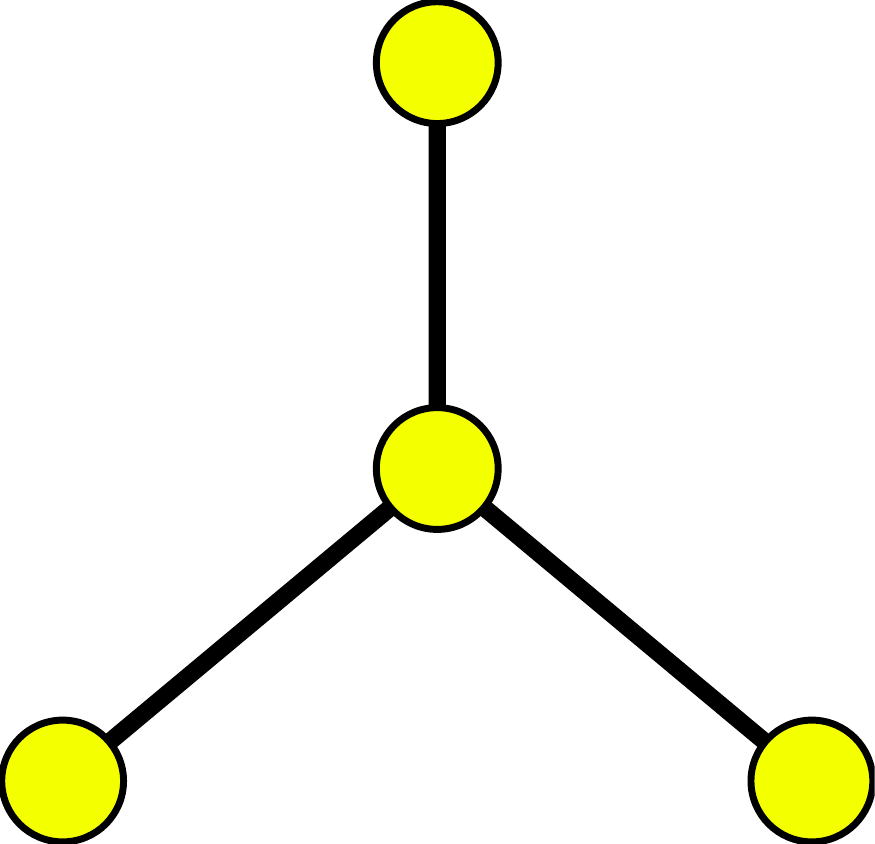}%
\subFig[3 matrix model, $\e=2$ (sphere) Ex 8]{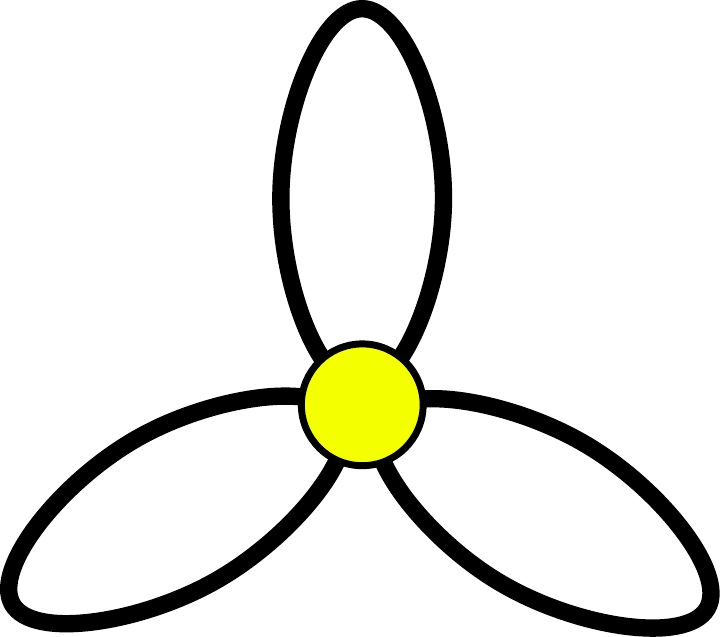}%
\subFig[3 matrix model, $\e=2$ (sphere) Ex 7]{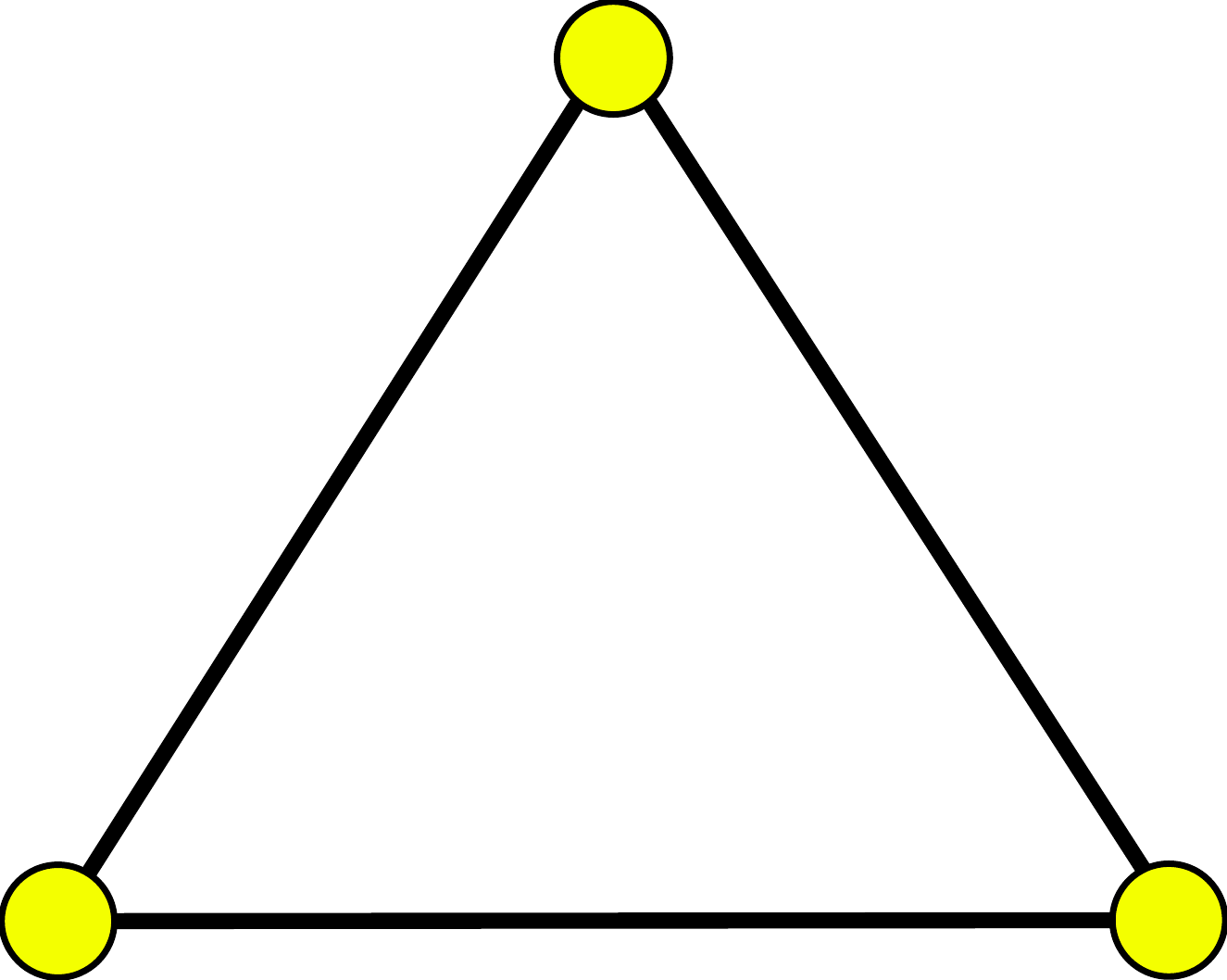}%
\subFig[5 matrix model, $\e=2$ (sphere) Ex 5]{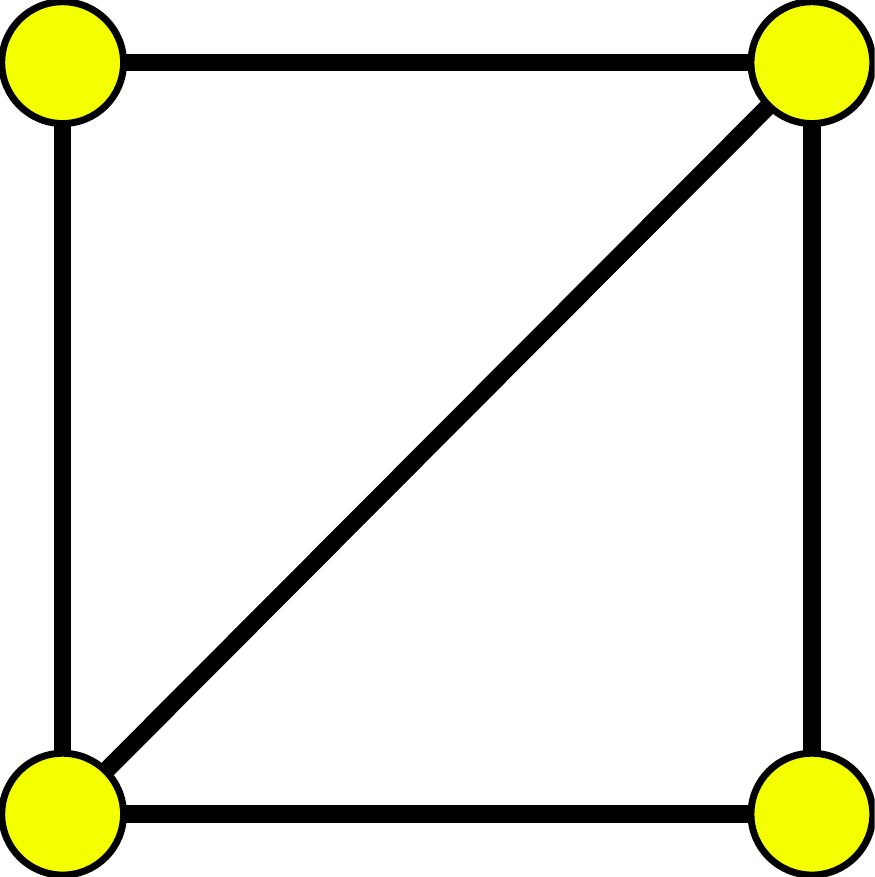}%
\subFig[4 matrix model, $\e=-2$ (pretzel) Ex 9]{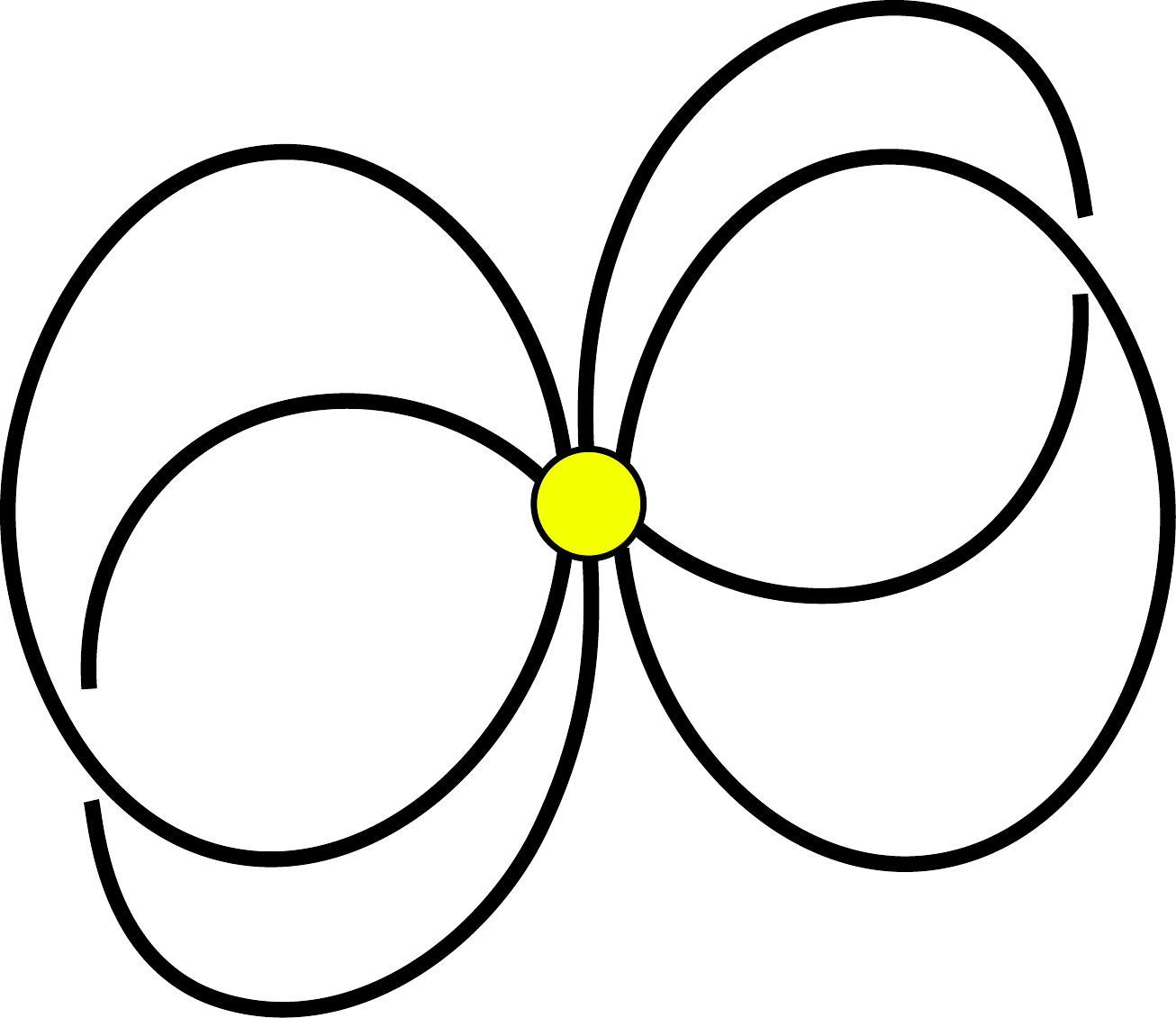}%
\end{arrangedFigure}

\section{Examples of matrix models}

$ \texttt{evfEVF}$

Recall that we introduced the function 
$\tau_r(n;\pb^{(1)},\pb^{(2)})$.
In what follows we use the conventions:
\be\label{convention}
\tau_r(\pb^{(1)},\pb^{(2)}):=\tau_r(0,\pb^{(1)},\pb^{(2)})=
\sum_\lambda r_\lambda s_\lambda(\pb^{(1)})s_\lambda(\pb^{(2)}),\quad 
r_\lambda=r_\lambda(0)
\ee

It depends on two sets $\pb^i=(p^{(i)}_1,p^{(i)}_2,\dots),\, i=1,2$ , as well as on the choice of 
an arbitrary function of the variable $r$.\footnote{
(This is an example of the so-called tau function, but we will not use this fact.}
As one of their sets, we will choose $\pb^2=\pb(X)$ like in (\ref{tau(nXp)}), and the second set
will be the set of arbitrary parameters. With $r=1$ we get 
\be\label{tau-vac}
\tau_1(\pb,X)=e^{\sum_{m>0}\frac{1}{m}p_m\ttr\left( X^m\right)}
\ee.
For example, if we take 
$$
r(x)=\frac{\prod_i^p(a_i+x)}{\prod_i^q(b_i+x)},
$$
and in addition $\pb^{1}=(1,0,0,\dots)$
we get the so-called hypergeometric function of the matrix argument:
\be\label{p-F-q}
{_pF}_q\left({a_1,\dots,a_p\atop b_1,\dots,b_q}\arrowvert X\right)=
\sum_\lambda \frac{{\rm dim}\,\lambda}{|\lambda|!}s_\lambda(X)
\frac{\prod_i^p(a_i+x)_\lambda}{\prod_i^q(b_i+x)_\lambda}
\ee 
Special cases: 
\be\label{e^tr}
e^{\ttr X}=\sum_\lambda s_\lambda(X)\frac{{\rm dim}\,\lambda}{|\lambda|!},\quad 
\ee
\be\label{det(1-X)}
\det(1-zX)^{-a} = \sum_\lambda z^{|\lambda|} (a)_\lambda s_\lambda(X)\frac{{\rm dim}\,\lambda}{|\lambda|!}
\ee

\paragraph{Integrals.}

We have

\begin{Theorem} Suppose $W_1,\dots, W_\f$ and $W_1^*,\dots, W_\V^*$ are dual sets.
Let sets $\pb^i=\left(p^{(i)}_1,p^{(i)}_2,p^{(i)}_3,\dots\right)$, $i=1,\dots,{\rm max}(\f,\V )$
be independent complex
parameters and $r{(i)},\,i=1,\dots,{\rm max}(\f,\V$ be a set of given functions in one variable.
\be\label{E-tau-X-p}
 E_{n_1,n_2}\left\{ \Dr_X\left[ 
 \prod_{i=1}^\f
 \tau_{r^{(i)}}\left(N;\pb^i, W_i \right)
\det \left(W_i \right)^{\alpha_i}
\right]\right\}  
\ee 
\be \label{Th1}
= \sum_\lambda \,r_\lambda\,
\hbar^{n_1\left(|\lambda|+\alpha N  \right)} 
\left(\frac{{\rm dim}\,\lambda}{|\lambda|!}\right)^{-n}
\prod_{i=1}^\f s_\lambda(\pb^i)
\prod_{i=1}^\texttt{v} s_{\lambda}\left(W_i^*\right)\det\left( W_i^* \right)^{\alpha},
\ee
where each $\tau_{r^{(i)}}\left(\pb^i, W_i \right)$ is defined by (\ref{tau(nXp)})
$$
r_\lambda=\left((N)_\lambda\right)^{-n_2}
\left( \frac{(N+\alpha)_\lambda}{(N)_\lambda} \right)^{n_1}
\prod_{i=1}^\f r^{(i)}_\lambda(n)
$$
Simirlaly
\be\label{E-tau-X-p*}
 E_{n_1,n_2}\left\{ \Dr_X\left[ \tau_{r^{(1)}}\left(N;\pb^1, W_1^* \right) \cdots
 \tau_{r^{(\texttt{v})}}\left(N;\pb^\texttt{v}, W_\texttt{v}^* \right)\right]\right\}  
\ee 
\be \label{Th1*}
= \sum_\lambda \,r_\lambda\,
\hbar^{n_1|\lambda|} 
\left(\frac{{\rm dim}\,\lambda}{|\lambda|!}\right)^{-n}
\prod_{i=1}^\f s_{\lambda}\left(W_i\right)
\prod_{i=1}^\texttt{v} s_\lambda(\pb^i),
\ee

\end{Theorem}
 
\br We recall the convention (\ref{convention})
 In (\ref{Th1}) $r_\lambda$ is the content product (\ref{content-product})
$$ 
r_\lambda=r_\lambda(0)=\prod_{(i,j)\in\lambda} r(j-i)
$$
 where
$$
r(x)=\left(N+x  \right)^{-n_2}\prod_{i=1}^\f r^{(i)}(x)
$$
\er 

To get examples we choose 

\begin{itemize}
 \item dual sets $W_1,\dots,W_\f\leftrightarrow W_1^*,\dots,W_\texttt{v}^*$
 \item the fraction of unitary matrices given by $n_2$
 \item the set of functions $r^{(i)},\,i=1,\dots,\f$
 \item the sets $\pb^{(i)},\,i=1,\dots,\f$
\end{itemize}

\br\label{simplifications} 
Important notice. Answers in some cases are further simplified. Let us mark two
cases

(i) Firstly, this is the case when 
the spectrum of the stars has the form 
\be\label{I-N-k}
{\rm Spect} \,W_i^* = {\rm Spect}\,\mathbb{I}_{N,k_i}=\diag \{1,1,\dots,1,0,0,\dots,0 \},\quad i=1,\dots,\texttt{v}
\ee
where $\mathbb{I}_{N,k_i}$ is the matrix
with $k_i$ units of the main diagonal. Such star monodromies are obtained in case source matrices have a rank
smaller than $N$.
Insertion of such matrices in the left hand sides of (\ref{Th1}) and (\ref{Th1*}) corresponds to the integration
over rectangular random matrices.
One should take into account
that
\[
 s_\lambda(\mathbb{I}_{N,k})=(k)_\lambda s_\lambda(\pb_\infty),
\]
where we recall the notation
\be\label{Pochhammer-lambda}
 (a)_\lambda:=(a)_{\lambda_1}(a-1)_{\lambda_2}\cdots (a-\ell+1)_{\lambda_\ell},
\ee

(ii) The case is the specification of the sets $\pb^i$, $i=1,\dots,\f$ according to the following

\bl\label{specializations} Denote 
\be\label{p_infty}
\pb_\infty =(1,0,0,\dots)
\ee
\be\label{p(a)}
\pb(a)=\left(a,a,a,\dots \right)
\ee
\be\label{p(t,q)}
\pb(\texttt{q},\texttt{t})=\left(p_1(\texttt{q},\texttt{t}),p_2(\texttt{q},\texttt{t}),\dots\right)\,,\quad
p_m(\texttt{q},\texttt{t})=  \frac{1-\texttt{q}^m}{1-\texttt{t}^m}
\ee 
Then
\be\label{Schur-t(a)}
\frac{s_\lambda(\pb(a))}{s_\lambda(\pb_\infty)}=(a)_\lambda\,,\quad \pb(a)=(a,a,a,\dots)
\ee
where $(a)_\lambda:=(a)_{\lambda_1}(a-1)_{\lambda_2}\cdots (a-\ell+1)_{\lambda_\ell}$, $(a)_n:=a(a+1)\cdots(a+n-1)$,
where $\lambda=(\lambda_1,\dots,\lambda_\ell)$ is a partition.
More generally
\be\label{Schur-t(t,q)}
\frac{s_\lambda(\pb(\texttt{q},\texttt{t}))}{s_\lambda(\pb(0,\texttt{t}))}=(\texttt{q};\texttt{t})_\lambda\,,
\ee
where $(\texttt{q};\texttt{t})_\lambda =
(\texttt{q};\texttt{t})_{\lambda_1}(\texttt{q t}^{-1};\texttt{t})_{\lambda_2}\cdots
(\texttt{q t}^{1-\ell};\texttt{t})_{\lambda_\ell}$ where
$(\texttt{q};\texttt{t})_k=(1-\texttt{q})(1-\texttt{q t})\cdots (1-\texttt{q t}^{n-1}) $
is $\texttt{t}$-deformed Pochhammer symbol. $(\texttt{q};\texttt{t})_0=1$ is implied.
\el

With such specifications, one can dinimish the number of the Schur functions in the right-hand side of 
(\ref{Th1}) (or of (\ref{Th1*})) and the right hand side can take one of the forms:
\be\label{form1}
\sum_\lambda r_\lambda s_\lambda(A)s_\lambda(B)
\ee
\be\label{form2}
\sum_\lambda r_\lambda s_\lambda(A)
\ee
\be\label{form3}
\sum_\lambda r_\lambda 
\ee
For (\ref{form1}) there is a determinant representation, for (\ref{form2}) there is a Pfaffian representation
and (\ref{form3}) can be rewritten as a sum of products.

For instance, one can take $r(x)=a+x$ and get
\be\label{(a)_lambda-poch)}
(a)_\lambda =\frac{\Gamma(a+\lambda_1-1)\Gamma(a+\lambda_2-2)\cdots \Gamma(a+\lambda_N-N)}
{\Gamma(a)\Gamma(a-1)\cdots \Gamma(a-N+1)}
\ee
Then we introduce $h_i=\lambda_i-i+N$ and write
\be\label{pochh-via-gamma}
\sum_\lambda\frac{(a)_\lambda}{(b)_\lambda}=
\sum_{h_1>\cdots >h_N\ge 0}\prod_{i=1}^N \frac{\Gamma(b-i+1)}{\Gamma(a-i+1)}\frac{\Gamma(h_i+a-N)}{\Gamma(h_i+b-N)}
\ee
where $\Gamma$ is the gamma-function. 
(In case the argument of gamma-function turns out to be a nonpositive integer one should keep in mind 
both the enumerator and denominator).
 See examples below.

\er

\bx\label{E1} See Example and Figure 2 (a). Take $X_1=Z$ and $r$ given by(\ref{r-rational}). 

(a) The example of (\ref{Th1}) can be chosen as follows
\be
E_{1,0}\left\{  
{_pF}_q\left({a_1,\dots,a_p\atop b_1,\dots,b_q}\arrowvert ZC_1\right)
{_{p'}F}_{q'}\left({a'_1,\dots,a'_{p'}\atop  b'_1,\dots,b'_{q'}}\arrowvert Z^\dag C_{-1}\right)
\det\left(Z Z^\dag  \right)^{\alpha}
\right\}=
\ee
\be
{_{p'}F}_{q'}\left({a_1,\dots,a_p,a'_1,\dots,a'_{p'},N+\alpha \atop b_1,\dots,b_q,
b'_1,\dots,b'_{q'},N}\arrowvert C_1 C_{-1}\right),
\ee
the corresponding determinantal representation see in (\ref{tau(nXp)}).

See  and Figure 2 (b) which is dual to (a) in this Figure.

(b) An example of (\ref{Th1*}) can be chosen as
$$
E_{1,0}\left\{
{_{p'}F}_{q'}\left({a_1,\dots,a_p,a'_1,\dots,a'_{p'},N+\alpha \atop b_1,\dots,b_q,
b'_1,\dots,b'_{q'},N}\arrowvert ZC_1 Z^\dag C_{-1}\right)\det\left(Z Z^\dag  \right)^{\beta}
\right\}=
$$
\be
\sum_\lambda s_\lambda(C_1)s_\lambda(C_{-1})
\frac{(N+\alpha)_\lambda(N+\beta)_\lambda)}{\left( (N)_\lambda \right)^2}
\frac{\prod_i^p(a_i)_\lambda}{\prod_i^q(b_i)_\lambda}\frac{\prod_i^{p'}(a_i)_\lambda}{\prod_i^{q'}(b_i)_\lambda},
\ee
The determinantal representation of the left hand side is given by (\ref{tau(nXY)}).
\ex

(c) Next example
\be
E_{0,1}\left\{e^{\sum_{m>0}\frac 1m p_m\ttr \left( UC_1U^\dag C_{-1} \right)^m }  \right\}
=
\sum_{\lambda} \frac{s_\lambda(\pb)s_\lambda(C_1)s_\lambda(C_{-1})}{s_\lambda(\mathbb{I}_N)}
\ee
It can be written as a determinant either if $\pb$ is specialized according to, then
it can be written in form, or if any of the matrices $C_1,C_{-1}$ has the same sprectrum
as $\mathbb{I}_{N.k}$ while the parameters $\pb$ are arbitrary. For instance
\be
E_{0,1}\left\{e^{\sum_{m>0}\frac 1m p_m\ttr \left( UC_1U^\dag \mathbb{I}_{N,k} \right)^m }  \right\}
=
\sum_{\lambda} s_\lambda(\pb)s_\lambda(C_1)\frac{(k)_\lambda}{(N)_\lambda}
\ee

(d) 
\be
E_{0,1}\left\{
e^{\sum_{m>0}\frac 1m \left(p_m^{(1)}\ttr \left( UC_1 \right)^m
+ p_m^{(2)}\ttr \left(U^\dag C_{-1}\right)^m  \right)}\det U^\alpha  
\right\}
=
\sum_{\lambda} s_\lambda(\pb^1)s_\lambda(\pb^2)s_\lambda(C_1C_{-1})\frac{(N+\alpha)_\lambda}{\left((N)_\lambda\right)^2}
\ee

\bx\label{E3} Graph 1 (c) yields
$$
E_{2,0}\left\{e^{\sum_{m>0} \frac 1m p^{(1)}_m \ttr\left(Z_1C_1)^m\right)+
\sum_{m>0} \frac 1m p^{(2)}_m \ttr\left(Z_2C_2)^m\right)}
\det\left(1-z Z_1^\dag C_{-1}Z_2^\dag C_{-2}\right)^{-a}\prod_{i=1}^3
\det\left(Z_i Z_i^\dag  \right)^{\alpha}
\right\}
$$
$$
=\sum_\lambda z^{|\lambda|} s_\lambda(\pb^1)s_\lambda(\pb^2) (a)_\lambda
\left(\frac{ (N+\alpha)_\lambda}{(N)_\lambda}\right)^3
$$
For a determinant representation see (\ref{tau(npp)}) 

\ex

\bx\label{E6} In the case below we use an open chain with $n$ edges as in Figures 1 (b), 
2 (b), 3 (a).
\be\label{Ex6-schur-1}
E_{n,0}\left\{
e^{\sum_{m>0}\frac 1m p^{(1)}_m \ttr\left((Z_1C_{1}Z_2C_2\cdots Z_nC_{n}Z_n^\dag C_{-n} 
\cdots Z_1^\dag C_{-1})^m\right)} \right\}=
\sum_{\lambda} s_\lambda(C_n) s_\lambda(C_{-1}) \frac{s_\lambda(\pb)}{s_\lambda(\pb_\infty)}
\prod_{i=1}^{n-1} \frac{s_\lambda(C_{i}C_{-i-1})}{s_\lambda(\pb_\infty)}
\ee
Graphs dual to the chain look like in Figure 3 (b).
$$
E_{n,0}\left\{e^{\sum_{m>0}\frac 1m p_m \ttr\left((Z_1^\dag C_{-1})^m\right)+
\sum_{m>0}\frac 1m p^{(n)}_m \ttr\left((Z_n C_{n})^m\right)}
\prod_{i=1}^{n-1}
e^{\sum_{m>0}\frac 1m p^{(i)}_m \ttr\left((Z_iC_{i}Z_{i+1}^\dag C_{-i-1})^m\right)}
\right\}
$$
\be\label{Ex6-schur-2}
=\sum_{\lambda} s_\lambda(\pb)s_\lambda(C_{1}C_2\cdots C_{n}C_{-n} \cdots C_{-1})
\prod_{i=1}^{n}\frac{s_\lambda(\pb^{i})}{s_\lambda(\pb_\infty)}
\ee

\ex

In case we specify each argument of the Schur functions in the right hand side  
except any two ones according to Remark 
\ref{simplifications}, we can write down a determinantal representation.

\bx\label{E7} Our graph is a polygon with $n$ edges and $n$ vertices (stars), see Figures 1 (b),
2 (a), 3 (c) and 4 (c) for examples.
\be
E_{n,0}\left\{
e^{\sum_{m>0}\frac 1m p^{(1)}_m 
\ttr\left( Z_1C_{1}Z_2C_2\cdots Z_nC_n  \right)^m
+
\sum_{m>0}\frac 1m p^{(2)}_m \ttr\left(Z_n^\dag
C_{-n}Z_{n-1}^\dag C_{1-n}\cdots Z_1^\dag C_{-1}\right)^m }
\prod_{i=1}^n
\det\left(Z_i Z_i^\dag  \right)^{\alpha}
\right\}  
\ee
$$
 =\sum_\lambda s_\lambda(\pb^1)s_\lambda(\pb^2)
 \prod_{i=1}^n\frac{(N+\alpha)_\lambda s_\lambda(C_{i}C_{-i-1})}{(N)_\lambda s_\lambda(\pb_\infty)} 
$$
(where we put $C_{-n-1}=C_{-1}$).

A graph dual to the polygon can be viewed as two-stars graph with $n$ edges which connect stars, see
Figures 1 (d) and 3 (d) as examples.
\be
E_{n,0}\left\{\prod_{i=1}^n e^{\sum_{m>0}\frac 1m p_m^{(i)} 
\ttr\left( Z_i C_{i}Z_{i+1}^\dag C_{-i-1}\right)^m}
  \right\} =  \sum_\lambda 
  s_\lambda(C_{1}C_2\cdots C_n) s_\lambda(C_{-n}C_{1-n}\cdots C_{-1})
  \prod_{i=1}^n \frac{s_\lambda(\pb^i)}{ s_\lambda(\pb_\infty)} 
\ee
To apply determinantal formalas one should use Remark \ref{simplifications}.

\ex

\bx\label{E8} Consider the star-graph with $n$-rays which end at other stars (see Figure 4 
(a) where $n=3$). 

\be
E_{n,0}\left\{e^{\sum_{m>0}\frac 1m p_m 
\ttr\left( Z_1C_{1}Z_1^\dag C_{-1}\cdots Z_nC_{n}Z_{n}^\dag C_{-n}\right)}\prod_{i=1}^n
\det\left(Z_i Z_i^\dag  \right)^{\alpha}
\right\}= 
\ee
$$
\sum_\lambda s_\lambda(\pb)s_\lambda(C_{-1}C_{-2}\cdots C_{-n})
\prod_{i=1}^n \frac{(N+\alpha)_\lambda s_\lambda(C_i) }{(N)_\lambda s_\lambda(\pb_\infty)}
$$
The similar model was studied in \cite{Chekhov-2014}, \cite{ChekhovAmbjorn}.
It has the determinantal representation (\ref{tau(nXp)}) in case all $W_i^*$ except one are of form
(\ref{I-N-k}). There is The determinantal representaion (\ref{tau(nXY)}) in case we specialize
the set $\pb$ according to Lemma \ref{specializations} and choose each $W_i^*$ except two
be in form (\ref{I-N-k}).

Now, let us choose the dual graph (this is petel graph. see (see Figure 4 (b) where $n=3$))
and consider
\be
E_{n,0}\left\{\det\left(1-z Z_1^\dag C_{-1}Z_2^\dag C_{-2}\cdots Z_n^\dag C_{-n}\right)^{-a}
\prod_{i=1}^n \det\left(Z_i Z_i^\dag  \right)^{\alpha} 
e^{\sum_{m>0}\frac 1m p_m^{(i)} \ttr\left((Z_iC_i)^m \right) }
\right\}=
\ee
$$
\sum_\lambda z^{|\lambda|}(a)_\lambda s_\lambda(\pb_\infty) s_\lambda(C_1C_{-1}C_2C_{-2}\cdots C_nC_{-n})
\prod_{i=1}^n \frac{(N+\alpha)_\lambda s_\lambda(\pb^i)}{(N)_\lambda s_\lambda(\pb_\infty)}
$$
By Remark \ref{simplifications} we find all cases where the determinantal representations 
(\ref{tau(npp)}) or (\ref{tau(nXp)}) exist.

\br\label{powers-symmetry} Notice the following symmetry: the left hand side produces the same
right hand side if we permute $\alpha \leftrightarrow a-N$.

\er

\ex


\begin{thebibliography}{99}



\bibitem{Alex} A.~Alexandrov,
``Intersection numbers on $\overline {\mathcal M}_{g,n}$ and BKP hierarchy'',
    JHEP 09 (2021) 013; DOI:
    10.1007/JHEP09(2021)013 ;
arXiv:2012.07573

\bibitem{Alex2} A. Alexandrov, ``KdV solves BKP'', 
Proc Natl Acad Sci U S A. 2021 Jun 22; 118(25):e2101917118.
doi: 10.1073/pnas.2101917118.
arXiv:2012.10448


\bibitem{ChekhovAmbjorn} J. Ambjorn and L. Chekhov \textit{The matrix model
for hypergeometric Hurwitz number},
Theoret. and Math. Phys., 1  
81:3 (2014), 1486-1498; arXiv:1409.3553


\bibitem{AOV} N. Amburg, A. Orlov, D. Vasiliev,  
``On Products of Random Matrices'', Entropy  vol 22, ussue 9,
https://doi.org/10.3390/e22090972

\bibitem{Bertola_} M. Bertola, M. I. Gekhtman, J. Szmigielski
``Strong asymptotics for Cauchy biorthogonal polynomials with application to the Cauchy two-matrix model''
Journal of Mathematical Physics 54(4) (2012)
    DOI:10.1063/1.4802455Corpus ID: 118796701
    
\bibitem{BrezinHikami} Brezin Hikami

\bibitem{Chekhov-2014}   Chekhov  (2014)  
    
    
\bibitem{DJKM1} E. Date, M. Jimbo, M. Kashiwara and T. Miwa, ``Transformation groups for soliton equations IV.
A new hierarchy of soliton equations of KP type'', {\it Physica} {\bf 4D},  343-365 (1982).

\bibitem{EOP} A. Eskin, A. Okounkov, R. Pandharipande,
``The theta characteristic of a branched covering'',
Adv. Math. {\bf 217} (2008) 873-888


\bibitem{GMMMO} A. Gerasimov, A. Marshakov, A. Mironov, A. Morozov, A. Orlov,
``Matrix models of two-dimensional gravity and Toda theory'', Nuclear Physics B 357 (2-3), 565-618 (1991)

\bibitem{H-review} J.Harnad, ``Weighted Hurwitz numbers and hypergeometric $\tau$-functions: an overview''
Proc. Symp. Pure Math. 93. 289-333 (2016) 

\bibitem{HO-Borel}  J. Harnad and A. Yu. Orlov, ``Matrix integrals as Borel sums of Schur function expansions'',
Symmetry and Perturbation Theory: Proc Int Conference SPT 2002, ed. S. Abenda, G. Gaeta, S. Walcher, pp.
116-123 World Scientific (2002);
arXiv nlin:0209035 

\bibitem{HLO} J. Harnad, J. W. van de Leur, A.Yu. Orlov, ``Multiple sums and integrals as neutral BKP tau functions'',
Theoretical and Mathematical Physics 168 (1) (2011) 951-962

\bibitem{paper1}  J. Harnad and A. Yu. Orlov,  
``Fermionic approach to bilinear expansions of Schur functions in Schur Q-functions'',
Proc. Amer. Math. Soc. 149, 4117-4131 (2021), 
arXiv:2008.13734 (2020).

\bibitem{paper2}   J. Harnad and A. Yu. Orlov, ``Bilinear expansions of lattices of KP $\tau$-functions in BKP
  $\tau$-functions: a fermionic approach'',  J. Math. Phys. 62, 013508 (2021);
  arXiv:22010.05055 (2020).

\bibitem{HO2005}   J. Harnad and A. Yu. Orlov, ``Fermionic construction of partition functions for 
two-matrix models and perturbative Schur function expansions'',
Journal of Physics A: Mathematical and General 39 (28), 8783 (2006); arxiv:0512056
      
\bibitem{HB}   J. Harnad and F. Balogh, ``Tau functions and their applications'',  Chapts. 5  and 7 and Appendix D,
  {\it  Monographs on Mathematical Physics}, Cambridge University Press  (in press, 2020).
    
\bibitem{Iv} V.N. Ivanov, ``Interpolation analogues of Schur Q-functions'',
{\it  Math. Sci.} {\bf  131}, 5495-5507, (2005).


\bibitem{Ivanov}  Vladimir N. Ivanov, ``The Dimension of Skew Shifted Young Diagrams, and Projective Characters of the 
Infinite Symmetric Group'', Journal of Mathematical Sciences (New York) 96 (1999) no.5 3517-3530,  
arXiv:math/0303169 
   
\bibitem{JM}  M. Jimbo and T. Miwa,  ``Solitons and infinite-dimensional Lie algebras'',
 {\it  Publ. Res. Inst. Math. Sci.}, {\bf 19} 943-1001 (1983).

\bibitem{KvdLbispec} V. Kac and J. van de Leur,
``The geometry of spinors and the multicomponent BKP and DKP hierarchies'',
CRM Proceedings and Lecture Notes {\bf 14}  (1998) 159--202

\bibitem{KvdLRoz}  V. G. Kac, N. Rozhkovskaya and J. van de Leur,
  ``Polynomial Tau-functions of the KP, BKP, and the s-Component KP Hierarchies'',
  J. Math. Phys. 62, 021702 (2021); https://doi.org/10.1063/5.0013017; 
  arXiv:2005.02665.
  
\bibitem{KazakovSW} V. Kazakov,  M. Staudacher and T. Wynter, ``Character Expension Methods 
for Matrix Models of Dually Weighted Graphs'', {\it Comm. Math. Phys.} {\bf 177} (1996) 451-468;
arxiv: hep-th/9502132
  
\bibitem{Kharchev} S. Kharchev, ``Kadomtsev-Petviashvili Hierarchy and Generalized Kontsevich Model'',  preprint
ITEP/TH-78/97, arXiv:hep-th/9810091

\bibitem{KMMMP} S. Kharchev, A. Marshakov, A. Mironov, A. Morozov, S. Pakuliak
``Conformal Matrix Models as an Alternative to Conventional Multi-Matrix Models'' , Nucl. Phys. B404 (1993)
717, arXiv:hep-th/9208044 

\bibitem{KMMOZ} S. Kharchev, A. Marshakov, A. Mironov, A. Orlov, A. Zabrodin,
``Matrix models among integrable theories: Forced hierarchies and operator formalism'',
Nuclear Physics B 366 (3), 569-601 (1991)
 
\bibitem{GKM} S. Kharchev, A. Marshakov, A. Mironov, A. Morozov, A. Zabrodin,
  ``Unification of all string models with $c<1$'',
  Phys. Lett.  B275  (1992) 311,

\bibitem{KMMM}  
S. Kharchev, A. Marshakov, A. Mironov and A. Morozov,
  ''Generalized Kazakov-Migdal-Kontsevich Model: group theory aspects'', 
  International Journal of Mod Phys A10 (1995) p.2015

\bibitem{Rangloom} Koch ,  Rangloom 
 
 
\bibitem{LZ} S. K. Lando, A. K. Zvonkin { Graphs on Surfaces and their Applications}, Encyclopaedia of Mathematical Sciences,
	Volume 141, with  appendix by D. Zagier, Springer, N.Y. (2004).
	
  
\bibitem{Lee2018} Junho Lee, A square root of Hurwitz numbers, Manuscripta Math. 162 (2020),
no. 1-2, 99-113; arxiv 1807.03631

\bibitem{L1} J. W. van de Leur, \textit{Matrix Integrals and Geometry of
Spinors}, { J. of Nonlinear Math. Phys.} {\bf  8},  pp. 288-311
(2001)

\bibitem{LO-2008}  J. W. van de Leur, A. Yu. Orlov, ``Random turn walk on a half line with creation 
of particles at the origin'', Physics Letters A 373 (31), 2675-2681 (2009) 
    
\bibitem{Mehta} Mehta, M.~L., {\it Random Matrices}, 3nd edition
(Elsevier, Academic, San Diego CA, 2004).
  
\bibitem{Mac} I.~G.~Macdonald, {\it Symmetric Functions and Hall Polynomials},
Clarendon Press, Oxford, (1995).

\bibitem{Sho} Sho Matsumoto, ``Alpha-Pfaffian, pfaffian point process and shifted Schur measure'',
Linear Algebra and its Applications 403 (2005) 369-398

\bibitem{Mir_thesis} A. Mironov, thesis

\bibitem{MMMZ} A. Mironov, V. Mishnyakov, A. Morozov, A. Zhabin, 
arXiv:2112.11371 

\bibitem{MM-genQ} A. Mironov, A. Morozov, ``Generalized Q-functions for GKM''
arXiv:2101.08759  [pdf, other]  hep-th
doi 10.1016/j.physletb.2021.136474

\bibitem{MMq} A.~Mironov and A.~Morozov,
``Superintegrability of Kontsevich matrix model'',
Eur. Phys. J. C (2021) 81: 270
https://doi.org/10.1140/epjc/s10052-021-09030-x
arXiv:2011.12917

\bibitem{MMZ}  A. Mironov, A. Morozov, A. Zhabin, ``Connection between cut-and-join and Casimir operators'',
Phys.Lett.B 822 (2021) 136668
arXiv:2105.10978,  doi
10.1016/j.physletb.2021.136668
 
\bibitem{MMZ-2} A. Mironov, A. Morozov, A. Zhabin, ``Spin Hurwitz theory and Miwa transform for the Schur Q-functions'',
arXiv:2111.05776  

\bibitem{MMN2019} A.D. Mironov, A. Yu. Morozov, S.M. Natanzon,
``Cut-and-join structure and integrability for spin Hurwitz numbers'',
 Eur. Phys. J. C 80 (2020) 97, arXiv:1904.11458

\bibitem{MMNO} A. D. Mironov, A. Yu Morozov, S. M. Natanzon, A. Yu Orlov, ``Around spin Hurwitz numbers'',
Letters in Mathematical Physics 111(5) (2021)
DOI: 10.1007/s11005-021-01457-3
arxiv:2012.09847 (2020)

\bibitem{MJD}  T. Miwa,  M. Jimbo and E. Date, ``Solitons. Differential equations, symmetries and infinite dimensional algebras''
 {\it Cambridge Tracts in Mathematics},  Cambridge University Press, Cambridge, U.K. (2000).

 \bibitem{Nim} J.~J.~C.~Nimmo, ``Hall-Littlewood symmetric functions and the BKP equation'',
 {\it J. Phys. A}, {\bf 23}, 751-60 (1990).

\bibitem{O-2004-New} A. Yu. Orlov, ``New solvable matrix integrals'',
International Journal of Modern Physics A 19 (supp02) (2004) 276-293

\bibitem{NO2020tmp} S. M. Natanzon, A. Yu. Orlov,  TMP (2020) 

\bibitem{OS2000} A. Yu. Orlov, D. M. Scherbin, ``Fermionic representation for basic
hypergeometric functions related to Schur polynomials''
arXiv:nlin/0001001 (2000)

\bibitem{OS-TMP} A. Orlov, D.M. Shcherbin,
{\it Hypergeometric solutions of soliton equations},
Theor.Math.Phys.
{\bf 128} (2001) 906-926; for the complete version see: A. Orlov, D.M. Shcherbin arXiv:nlin/0001001

\bibitem{Or}  A.~Yu.~Orlov, ``Hypergeometric Functions Related to Schur Q-Polynomials and the BKP Equation''.
 {\it Theor. Math. Phys.} {\bf 137} (2), 1574-1589 (2003)

\bibitem{ONimmo} J. J. C. Nimmo, A. Yu. Orlov, ``A relationship between rational and multi-soliton 
solutions of the BKP hierarchy'', Glasgow Mathematical Journal 47 (A), (2005) 149-168

\bibitem{ShiotoOrlov} A.Yu.Orlov, T.Shiota, ``Schur function expansion for normal matrix model 
and associated discrete matrix models'', Physics Letters A 343 (5), 384-396 (2005)

\bibitem{OST-I}  A.~Yu.~Orlov, T. Shiota, K. Takasaki, ``Pfaffian structures and certain solutions to BKP hierarchies I. 
Sums over partitions'', arxiv:1201.4518

\bibitem{OST-II}  A.~Yu.~Orlov, T. Shiota, K. Takasaki, ``Pfaffian structures and certain solutions to BKP hierarchies II. 
Multiple integrals'', arxiv:arXiv:1611.02244
 
\bibitem{Sato} M.~Sato. ``Soliton equations as dynamical systems on infinite dimensional Grassmann manifold''
{\it Kokyuroku, RIMS} 30-46, (1981).

\bibitem{Serg} A. Sergeev, The tensor algebra of the identity representation as a module over the
Lie superalgebras Gl.n;m/ and Q(n),
Math. Sb. USSR, {\bf 51} (1985) 419-427

\bibitem{Strembridge} Strembridge

\bibitem{Takasaki} K. Takasaki,  ``Initial value problem for the Toda lattice hierarchy'',
 {\it Adv. Stud. Pure Math.} {\bf 4}, 139-163 (1984).

\bibitem{TakashiLMP} T. Takebe, ``Representation Theoretical meaning of the Initial Value Problem
for the Toda Lattice Hierarchy'', Letters in Math. Phys. {\bf 21}: 77-84 (1991)

\bibitem{Takasaki-95} K.Takasaki,
``Toda Lattice Hierarchy and Generalized String Equations'',  Commun.Math.Phys. 181 (1996) 131

\bibitem{TW} C. A. Tracy, H. Widom, ``A Limit Theorem for Shifted Schur Measures'',
Duke Mathematical Journal 123 (2004), 171-208

\bibitem{UT} K.Ueno and K.Takasaki, \textit{Toda lattice hierarchy}, {\it Adv. Stud. Pure Math.} {\bf  4},  1-95 (1984).
	
\bibitem{VK}	 Vilenkin  Klimyk  ; Richardson
	
\bibitem{Serbian} M. Vuletic, ``Schifted Schur Process and Asymptotics of Large Random
Strict Plane Partitions'', 
Mathematics, Physics
International Mathematics Research Notices,     DOI:10.1093/IMRN/RNM043Corpus ID: 14774050,
arXiv:math-ph/0702068v1	
	
\bibitem{You} Y.~You,  ``Polynomial solutions of the BKP hierarchy and projective representations of symmetric groups'',
 in: {\it Infinite-Dimensional Lie Algebras and Groups},
 {\it Adv. Ser. Math. Phys.} {\bf 7} , World Sci. Publ., Teaneck, NJ (1989).

\bibitem{ZinZub}  P. Zinn-Justin, J. B. Zuber, ``On some integrals over the $U(N)$ unitary group and 
their large $N$ limit'',   J.Phys. A36:3173-3194 (2003) 
 
\bibitem{Zuber} J.B.Zuber, ``On the large $N$ limit of matrix integrals over the orthogonal group'',
arxiv:0805.0315
 
 
 
\end{thebibliography}
\end{document}